\documentclass[aps,prd,twocolumn,showpacs,nofootinbib,eqsecnum,superscriptaddress]{revtex4-1}

\usepackage[centertags]{amsmath}
\usepackage{amssymb}
\usepackage{latexsym}
\usepackage{enumerate}
\usepackage{graphicx}
\usepackage{mathrsfs}
\usepackage[hidelinks]{hyperref}
\usepackage{stmaryrd}
\usepackage{import}
\usepackage{tensor}
\usepackage{xcolor}
\usepackage{bm}
\usepackage{multirow}
\usepackage{breakurl}
\usepackage{float}
\usepackage{lipsum}
\usepackage{siunitx}
\usepackage{comment}
\usepackage{mathtools}
\usepackage{autobreak}
\usepackage[caption=false]{subfig}

\mathtoolsset{centercolon}

\usepackage{stmaryrd}
\usepackage{color}
\usepackage{amsmath}
\usepackage{amssymb}
\usepackage{latexsym}
\usepackage{enumerate}
\usepackage{graphicx}
\usepackage{mathrsfs}
\usepackage{physics}
\usepackage{tabularx}
\usepackage{mathtools}

\newcommand{\be}{\begin{equation}}
\newcommand{\ee}{\end{equation}}
\newcommand{\bea}{\begin{eqnarray}}
\newcommand{\eea}{\end{eqnarray}}
\newcommand{\ba}{\begin{array}}
\newcommand{\ea}{\end{array}}
\newcommand{\bi}{\begin{itemize}}
\newcommand{\ei}{\end{itemize}}

\newcommand{\D}{\Delta}

\newcommand{\la}{\lambda}
\newcommand{\vp}{\varphi}
\renewcommand{\O}{\Omega}

\renewcommand{\th}{\theta}

\newcommand{\rp}{r}
\newcommand{\thetap}{\theta}
\newcommand{\phip}{\varphi}
\newcommand{\tp}{t}

\newcommand{\rmax}{r_\mathrm{max}}
\newcommand{\rmin}{r_\mathrm{min}}
\newcommand{\thetamin}{\theta_\mathrm{min}}
\newcommand{\thetamax}{\theta_\mathrm{max}}

\newcommand{\cE}{\mathcal{E}}
\newcommand{\cLz}{\mathcal{L}_z}
\newcommand{\cQ}{\mathcal{Q}}

\newcommand{\cS}{\mathcal{S}}
\newcommand{\cO}{\mathcal{O}}

\newcommand{\cH}{\mathcal{H}}

\makeindex

\begin{document}

\title{\texorpdfstring{Post-Newtonian expansion of gravitational energy and angular momentum fluxes:\\
inclined spherical orbits about a Kerr black hole}{Post-Newtonian expansion of energy and angular momentum fluxes:
inclined spherical orbits about a Kerr black hole}}

\begin{abstract}
    We present analytical expressions for the fluxes of energy and angular momentum from a point mass on an inclined spherical orbit about a Kerr black hole. The expressions are obtained using the method of Mano, Suzuki and Takasugi to construct analytical solutions of the Teukolsky equation, and are given as post-Newtonian expansions valid through 12PN, with arbitrary values for the inclination parameter $x$ and black hole spin $a$. We characterize the structure of the PN expansions in terms of their dependence on $x$ and $a$, and we validate our results against numerical calculations.
\end{abstract}
    
\author{Jezreel C. Castillo}
\affiliation{Department of Physics and Astronomy, University of North Carolina at Chapel Hill, Chapel Hill, North Carolina 27599}
\author{Charles R. Evans}
\affiliation{Department of Physics and Astronomy, University of North Carolina at Chapel Hill, Chapel Hill, North Carolina 27599}
\affiliation{School of Mathematics and Statistics, University College Dublin, Belfield D04 N2E5, Dublin 4, Ireland}
\author{Chris Kavanagh}
\affiliation{School of Mathematics and Statistics, University College Dublin, Belfield D04 N2E5, Dublin 4, Ireland}
\author{Jakob Neef}
\affiliation{School of Mathematics and Statistics, University College Dublin, Belfield D04 N2E5, Dublin 4, Ireland}
\author{Adrian Ottewill}
\affiliation{School of Mathematics and Statistics, University College Dublin, Belfield D04 N2E5, Dublin 4, Ireland}
\author{Barry Wardell}
\affiliation{School of Mathematics and Statistics, University College Dublin, Belfield D04 N2E5, Dublin 4, Ireland}

\maketitle

\section{Introduction}

With the recent adoption of the Laser Interferometer Space Antenna (LISA) mission by the European Space Agency (ESA), there is a growing urgency to build gravitational wave models of the expected signal from LISA sources \cite{AmarETC07}. Sources of particular interest are extreme-mass-ratio inspirals (EMRIs), binary systems involving a compact object (the \textit{secondary}) of mass $\mu$ in orbit about a massive central black hole (the \textit{primary}) of mass $M$, with mass ratios of about $\mu/M \sim 10^{-4}-10^{-6}$. Unlike sources detected by the LIGO-Virgo-KAGRA Collaboration (LVK) \cite{AbboETC16a,AbboETC17,LIGOScientific:2020ibl,KAGRA:2021vkt}, EMRI sources are expected to be highly eccentric, inclined and long-lasting, allowing for high precision parameter estimation \cite{GairETC17}.

Modelling EMRI sources is best done using techniques from black hole perturbation theory, in which $\mu/M \ll 1$ acts as a small expansion parameter. The overarching programme, known as self-force (SF) theory \cite{Bara09,PoisPounVega11} models the EMRI as a compact object causing a perturbation of a background black hole spacetime. The self-force programme has recently produced its first post-adiabatic (PA) waveforms \cite{WardETC23} suitable for LISA data analysis \cite{Burke:2023lno}. These initial results were restricted to quasicircular inspirals without spin, and work is now underway towards producing similar models for generic (eccentric and inclined/precessing) EMRIs.   

The incorporation of eccentricity and inclination in EMRI models is essential. For comparable-mass binaries such as those observed by LVK, eccentricity is efficiently radiated away through the process of circularization \cite{Pete64}. This is not the case for EMRIs, which are expected to settle at some moderate eccentricity \cite{GlamKenn02}. Furthermore, there is no equivalent ``equatorialization'' process that applies to inclination, and, in fact, numerical calculations of the evolution of binary systems have shown mild increases in the inclination as the system progresses through the inspiral \cite{Hugh00b,DellETC24}.

The work presented here is part of a larger effort to incorporate eccentricity and inclination into analytical expressions for self-force quantities. The general approach to producing analytical expressions relies on a double expansion of the Einstein equations:
\begin{enumerate}
    \item A \textit{self-force} expansion, which recasts the problem in terms of solutions of the Teukolsky equation for perturbations of Kerr spacetime;
    \item A \textit{post-Newtonian} expansion of the self-force equations of motion.
\end{enumerate}
As shown by Mano, Suzuki, and Takasugi (MST) \cite{ManoSuzuTaka96a,ManoSuzuTaka96b}, the Teukolsky equation admits analytical solutions in terms of an infinite, uniformly convergent sum of special functions. At any given order in a post-Newtonian (PN) expansion the sum is in fact finite, and the method yields closed-form analytical solutions.

To date, this programme has predominantly focused on binary configurations in which the secondary's orbit is in the equatorial plane of the primary's spacetime. Orbital inclination has largely been ignored, the notable exception being the calculation of expressions for the asymptotic flux of radiation from inclined, eccentric orbits up to 5PN \cite{IsoyETC22}. This \textit{dissipative} calculation only required solutions to the Teukolsky equation that are valid infinitely far from the source. There have not yet been any attempts at incorporating inclination into analytical calculations of \textit{conservative} quantities, which require solutions to the Teukolsky equation on the worldline of the secondary. Part of the difficulty in doing so is dealing with angular functions; for equatorial orbits, the angular functions are merely constants \cite{KavaOtteWard16,Munn23} that can be evaluated separately, while for inclined orbits the angular functions are parametrized by the motion along the polar angle. The exact structure of the angular functions then needs to be examined before any analytical expression can be constructed.

In this paper, we take a next step in the programme, and compute the fluxes of gravitational-wave energy and angular momentum from a test mass on an inclined spherical orbit about a Kerr black hole. We show that by neglecting eccentricity we can push the calculation to high PN order relatively easily, and we give explicit results up to 12PN. The resulting PN-SF expansions are exact functions of the inclination $x$ and black hole spin $a$, without any further expansions or approximations. In addition to generating novel dissipative results, we also anticipate that many of the techniques developed in this work will be immediately applicable to calculations of conservative quantities, although we leave the actual calculation of those quantities to future work.

The paper is organized as follows. In Sec. II we give a brief discussion of geodesic orbits about a Kerr black hole, then specializing to spherical inclined orbits. In Sec. III, an overview of the gravitational field generated by a point mass is given, where we discuss the following techniques used to expand the gravitational fields in a PN series, first providing a brief description of the MST method used for the radial function, then a discussion on the functional dependence of the angular functions in our chosen parameterization. Afterwards we briefly discuss our truncation scheme for the formally infinite sum over modes. In Sec. IV, we present our results for the PN expanded gravitational fluxes at infinity, starting with the overall structure, proceeding to noteworthy aspects in the individual flux components, and then to a comparison with a numerical calculation. Corresponding expansions for the flux at the horizon are given in Appendix \ref{sec:horizon}. In Appendix \ref{sec:retrograde}, we briefly discuss the compatibility of retrograde orbits with our flux expansions, providing a side-by-side comparison with numerical flux results for prograde and retrograde orbits. Higher order expansions, valid to 12PN, are available through the \texttt{PostNewtonianSelfForce} package of the Black Hole Perturbation Toolkit \cite{PostNewtonianSelfForce, BHPTK18} and as an added repository \cite{UNCGrav22}. Throughout the paper we work with a $(-,+,+,+)$ metric signature and we use geometrized units such that $G = 1 = c$.

\section{Spherical orbits in Kerr spacetime} \label{sec:geodesics}

\subsection{Timelike geodesic orbits}
In Boyer-Lindquist coordinates
$(t,r,\th,\vp)$ the spacetime of a Kerr black hole of mass $M$ and spin $a$ is defined by the line element
\begin{multline}
\label{eqn:kerrLineElement}
ds^2 = -\left(1-\frac{2Mr}{\Sigma}\right) dt^2 +\frac{\Sigma}{\Delta} dr^2
- \frac{4Mar\sin^2\theta}{\Sigma}dt d\varphi \\
+\Sigma d\theta^2+\frac{\sin^2\theta}{\Sigma}\left(\varpi^4-a^2\Delta \sin^2\theta \right) d\varphi^2 ,
\end{multline}
where $\Sigma =r^2+a^2\cos^2\theta$, $\Delta =r^2-2Mr+a^2$, and $\varpi = \sqrt{r^2 + a^2}$.

Timelike geodesics of Kerr spacetime admit three constants of motion:
the specific energy $\cE$, the $z$-component of the specific
angular momentum $\cLz$, and the Carter constant $\cQ$.
The specific energy and angular momentum are given by the projection of the four-velocity along the timelike and azimuthal Killing vectors of the spacetime,
\begin{align}
\cE &= -\xi^\mu_{(t)}u_\mu = -u_t , \\
\cLz &= \xi^\mu_{(\varphi)}u_\mu=u_\varphi,
\end{align}
while the Carter constant is obtained as a projection with the Killing tensor,
\begin{equation}
\cQ = K^{\mu\nu}u_\mu u_\nu - (\cLz-a\cE)^2.
\end{equation}

Inverting these relations, we then obtain the geodesic equations as a decoupled system of first-order ordinary differential equations \cite{Cart68,Schm02,DrasHugh06,Mino03,FujiHiki09}:
\begin{subequations}
\begin{align}
    \left(\frac{d\rp}{d\lambda} \right)^2 & = R(\rp) , \label{eq:R} \\
    \left(\frac{d\thetap}{d\lambda}\right)^2 &= \Theta(\thetap) ,  \label{eq:Th}\\
    \frac{d\phip}{d\lambda} &= \Phi^{(r)}(\rp)+\Phi^{(\theta)}(\thetap) -a\cE, \label{eq:Ph}\\
    \frac{d\tp}{d\lambda} &= T^{(r)}(\rp)+T^{(\theta)}(\thetap) +a\cLz, \label{eq:T}
\end{align}
\end{subequations}
where
\begin{subequations}
\begin{align}
    \label{eqn:PhiR}
    R(r) &= \left[\cE \varpi^2 - a\cLz \right]^2
    -\Delta \left[r^2+(\cLz-a \cE )^2+\cQ\right], \\
    \Theta(\theta) &= \cQ-\cLz^2 \cot^2 \theta-a^2(1-\cE ^2) \cos^2\theta, \\
    \Phi^{(r)}(r) &= a \cE \frac{\varpi^2}{\D}
    -\frac{a^2\cLz}{\Delta}, \\
    \label{eqn:PhiTh}
    \Phi^{(\theta)}(\theta) &= \cLz \csc^2\theta, \\
    \label{eqn:TR}
    T^{(r)}(r) &= \cE \frac{\varpi^4}{\Delta}
    - a \cLz\frac{\varpi^2}{\Delta}, \\
    \label{eqn:TTh}
    T^{(\theta)}(\theta) &= -a^2 \cE\sin^2\theta.
\end{align}
\end{subequations}
Here, we have introduced the Mino time parameter, $\lambda$, which is defined in terms of proper time, $\tau$, by $d\tau = \Sigma d\lambda$.

Any timelike geodesic is fully characterized by the conserved quantities $\{\cE,\cLz,\cQ\}$. Focusing on bound orbits and in analogy to Keplerian mechanics, it is convenient to also introduce the alternative parameter set $\{p,e,x\}$, representing the relativistic equivalent of the semi-latus rectum
\begin{equation}
    p= \frac{2 \rmax\rmin}{M(\rmax+\rmin)},
\end{equation}
and eccentricity
\begin{equation}
    e= \frac{\rmax - \rmin}{\rmax + \rmin},
\end{equation}
along with the inclination parameter
\begin{equation}
    x = \sin\thetamin.
\end{equation}
Here, $\rmin$ and $\rmax$ are the radial turning points of the motion and $\th_\mathrm{min}$ is the polar turning point\footnote{Our choice of inclination parameter $x$ differs from the parameter $\iota$ that is also in common use \cite{Hugh01}, defined as $\cos\iota = \mathcal{L}_z^2/\sqrt{\mathcal{L}_z^2+\mathcal{Q}}$.}. A bijective relationship between the two parameter sets $\{\cE, \cLz, \cQ\}\leftrightarrow\{p,e,x\}$ can be derived by imposing $R(\rmin) = R(\rmax)= \Theta(\thetamin) = 0$ (or $R(p) = R'(p)= \Theta(\thetamin) = 0$ in the case $e=0$) \cite{Schm02}.

We can next parametrise the position along a given orbit using modified versions of the Darwin orbital phase parameters \cite{Darw59,Darw61,DrasHugh04},
\begin{align}
    \rp(\chi_r) &= \frac{pM}{1+e \cos \chi_r}, \label{eq:rfunc} \\
    \cos\thetap(\chi_\theta) &= \sqrt{1-x^2}\cos\chi_\theta. \label{eq:thfunc}
\end{align}

\subsection{Spherical orbits and post-Newtonian expansions}

We now consider the special case $e=0$, in which the orbit is \textit{spherical} with constant radius $\rp = p M$, though in general still inclined and precessing \cite{Warb15}. A natural candidate for a post-Newtonian expansion parameter is then $1/p$, which is small for large orbital separations and correspondingly low velocities.

It is now straightforward to obtain series solutions for $\{\cE(p,x), \cLz(p,x), \cQ(p,x)\}$ using the equations $R(p) = R'(p) = \Theta(\thetamin) = 0$.  We find the first few orders are:
\begin{align}
    \cE&(p,x) = 1 - \frac{1}{2p} + \frac{3}{8 p^2} - \frac{a x}{p^{5/2}} + \left[\frac{27}{16} + \frac{a^2 x^2}{2} \right] \frac{1}{p^3} \nonumber \\
     & - \frac{9ax}{2 p^{7/2}} + \left[\frac{675}{128} + a^2 \left(\frac{23 x^2}{4}-2\right) \right]\frac{1}{p^4}+ \mathcal{O}\left(\frac{1}{p^{9/2}}\right) ,\\
    \cLz&(p,x) = x p^{1/2}\left[1 + \frac{3}{2p}-\frac{3 a x}{p^{3/2}} + \frac{27- 4 a^2 (1 - 3 x^2)}{8 p^2}\right. \nonumber \\
    & \left.-\frac{15 a x}{2p^{5/2}} + \left[\frac{135 + 4 a^2 (31 x^2-11)}{16}\right] \frac{1}{p^3}+ \mathcal{O}\left(\frac{1}{p^{7/2}}\right)\right],\\
    \cQ&(p,x) = (1-x^2) p\left[ 1+\frac{3}{p}-\frac{6 a x}{p^{3/2}}+\frac{9+3 a^2 x^2}{p^2}\right. \nonumber \\
    & \left.-\frac{24 a x}{p^{5/2}}+\frac{27+a^2 (29 x^2-8)}{p^3}+ \mathcal{O}\left(\frac{1}{p^{7/2}}\right)\right].
\end{align}
Note that these are power series in $1/p$, but at any given order in $1/p$ they are exact functions of $a$ and $x$. Higher orders expressions in $1/p$ can be found at \cite{BHPTK18,UNCGrav22}. 

Our choice of parameterising the polar motion with $\chi_\theta \equiv \chi$ requires us to express the Mino time parameter $\lambda$ as a function of $\chi$. Using Eqs.~\eqref{eq:Th} and \eqref{eq:thfunc} we obtain a differential equation relating the two \cite{FujiHiki09},
\begin{align}
    \frac{d\chi}{d\lambda} = \sqrt{a^2(1-\cE^2)(z_+-z_-\cos^2\chi)}, \label{eq:polarchi}
\end{align}
where $z_- = (1-x^2)$ and $z_+ = \cQ/(a^2(1-\cE^2)(1-x^2))$ are the roots of $\Theta(\thetap)$.
Since we are considering bound orbits, the motion can also be described in terms of a discrete spectrum of frequencies. For a spherical inclined orbit, we only need to consider the (Mino-time) polar, azimuthal, and coordinate time frequencies $\{\Upsilon_\theta,\Upsilon_\varphi,\Upsilon_t\}$ \cite{FujiHiki09}.
In terms of Mino time the coordinate functions are
\begin{align}
    \label{eqn:tpOfLa}
    \tp(\lambda) &= \Upsilon_t  \lambda + \Delta t^{(\th)}(\la) ,
    \\
    \label{eqn:phipOfLa}
    \phip(\lambda) &= \Upsilon_{\varphi} \lambda + \Delta \phip^{(\th)}(\la) ,
\end{align}
where apart from the linear in $\lambda$ portion, the remaining pieces are oscillatory functions of $\lambda$. Expressions for $\Upsilon_t$ and $\Upsilon_\varphi$ are then given by
\begin{align}
    \Upsilon_t &= T^{(r)}(p)+a\cLz+\left\langle T^{(\th)}(\th_p)\right\rangle_\la, \\
    \Upsilon_\varphi &= \Phi^{(r)}(p) - a\cE + \left\langle \Psi^{(\th)}(\th_p)\right\rangle_\la,
\end{align}
where angle brackets denote a time average,
\begin{align}
    \left\langle F(\lambda) \right\rangle_\lambda = \frac{1}{\Lambda}\int_{0}^{\Lambda}F(\lambda)d\lambda,
\end{align}
while $\Upsilon_\theta$ is obtained by integrating Eq. \eqref{eq:polarchi} 
\begin{align}
    \frac{\Upsilon_\theta}{2\pi} = \left(\int_{0}^{2\pi} \frac{d\chi}{\sqrt{a^2(1-\mathcal{E}^2)(z_+-z_-\cos^2\chi)}}\right)^{-1}.
\end{align}
These expressions are amenable to a straightforward PN expansion.  More importantly, the fundamental frequencies with respect to coordinate time, which are given by 
\begin{align}
    \Omega_\theta = \frac{\Upsilon_\theta}{\Upsilon_t}, \qquad \O_\vp = \frac{\Upsilon_\varphi}{\Upsilon_t} ,
\end{align}
can likewise be PN expanded to arbitrarily high order. The first few orders of the polar frequency are
\begin{align}
    \Omega_{\theta} = \frac{1}{p^{3/2}} &\left[ 1- \frac{3ax}{p^{3/2}}-\frac{3a^2(1-3x^2)}{4p^2} -\frac{a^2(2-3x^2)}{p^3}\right. 
    \nonumber \\
    &\left. +\frac{3a^3x(9-7x^2)}{4p^{7/2}}+O\left(\frac{1}{p^4}\right)\right] .
\end{align}  
In our convention, the sign of $\cLz$ and $x$ flip to negative for retrograde orbits, which implies that $\Omega_{\phip}$ 
does as well for wide orbits.  We find the leading expansion of $\Omega_{\phip}$ to be
\begin{align}
\label{eq:AzO}
    \Omega_{\phip} &= \frac{1}{p^{3/2}}\left[\text{sgn}\, x \left(1 
    -\frac{a^2(2-3x^2)}{p^3} - \frac{3a^2(1-3x^2)}{4p^{2}}\right)
    \right.
    \nonumber \\
    &\left.+\frac{a(2-3|x|)}{p^{3/2}} - \frac{3a    ^2 x}{2 p^{2}} 
     - \frac{3a^3(4-9|x|-2x^2+7|x|^3)}{4p^{7/2}} \right.
     \nonumber \\ 
     &\left. + O\left(\frac{1}{p^4}\right)\right].
\end{align}
Finally, it is worth pointing out that in the special case of polar orbits ($x = 0$), the rate of azimuthal advance 
reduces to
\begin{align}
\label{eq:AzOpol}
    \Omega^{\rm pol}_{\phip} = \frac{2 a}{p^3} 
    \left(1 - \frac{3 a^2}{2 p^2} - \frac{a^2}{p^3} + \frac{27 a^4}{16 p^4} +O\left(\frac{1}{p^5}\right)\right) ,
\end{align}
and is solely due to frame dragging ($a \ne 0$).  The angular rate is seen to be lower by 1.5 PN orders and leads to a 
special case in 
the radiated fluxes.  The frequently used alternative PN compactness parameter, $y = (M\Omega_{\phip})^{2/3}$, may now be readily derived from Eq.~\eqref{eq:AzO} (or Eq.~\eqref{eq:AzOpol}) in terms of $1/p$, or vice versa.

\section{Perturbations of Kerr spacetime} \label{sec:scalars}

\subsection{Teukolsky equation}

Gravitational perturbations of Kerr spacetime can be represented in terms of the Weyl scalar $\Psi_4$, which satisfies the spin-weight $s=-2$ Teukolsky equation \cite{Teuk72,Teuk73},
\begin{align}
    \cO_4 \Psi_4 = 8\pi \cS_4 T.
\end{align}
Here $\cO_4$ is the Teukolsky operator and $\cS_4$ is the corresponding decoupling operator that acts on the stress-energy $T_{\mu \nu}$ (see appendix A of \cite{Wardell:2024yoi}).
Working in the Kinnersley tetrad, the Teukolsky equation admits solutions in terms of separation of variables using the ansatz
\begin{widetext}
\begin{align}
 \zeta^{4} \Psi_4 = \int_{-\infty}^\infty \sum_{\ell m} {}_{-2}\psi_{\ell m\omega}(r){}_{-2}S_{\ell m}(\theta, \varphi; a \omega)e^{-i\omega t} d\omega, \label{eq:psiDef} 
\end{align}
where $\zeta = r-i a \cos\theta$. A similar separation ansatz is also used for the source term,
\begin{align}
  8 \pi \zeta^{4}\cS_4 T = 
    -\frac{1}{2 \Sigma} \int_{-\infty}^\infty \sum_{\ell m} {}_{-2}T_{\ell m\omega}(r){}_{-2}S_{\ell m}(\theta, \varphi; a \omega)e^{-i\omega t} d\omega. \label{eq:TDef}
\end{align}

The functions ${}_{s} S_{\ell m}(\theta, \phip; a \omega)$ and ${}_{s} \psi_{\ell m \omega}(r)$ satisfy the spin-weighted spheroidal harmonic and Teukolsky radial equations, respectively,
\begin{equation}
  \label{eq:SWSH}
  \bigg[\dfrac{d}{dz} \bigg((1-z^2)\dfrac{d}{dz} \bigg)
  +a^2 \omega^2 z^2 -\frac{(m+s z)^2}{1-z^2} - 2 a s \omega z +s + A\bigg] {}_{s} S_{\ell m} = 0,
\end{equation}
\begin{equation}
  \label{eq:radTeukEq}
  \bigg[\Delta^{-s} \dfrac{d}{dr}\bigg( \Delta^{s+1}\dfrac{d }{dr}\bigg)
  +\frac{K^2 - 2 i s (r-M)K}{\Delta} + 4 i s \omega r - {}_s \lambda_{\ell m} \bigg]{}_{s} \psi_{\ell m \omega} = {}_{s} T_{\ell m \omega},
\end{equation}
\end{widetext}
where $z := \cos \theta$, $A:= {}_s \lambda_{\ell m}+2 a m \omega -a^2 \omega^2$ and $K:=(r^2+a^2)\omega-a m$.  The eigenvalue ${}_s \lambda_{\ell m}$ depends on the value of $a\omega$, and we take $s=-2$ to obtain the Teukolsky equation for $\Psi_4$.
When considering homogeneous solutions to these equations we use standard normalization conventions such that the spin-weighted spheroidal harmonics are unit-normalized over the sphere (similar to the spherical harmonics \cite{AbraSteg72}) and the Teukolsky radial functions have unit transmission coefficients.

In the context of perturbations sourced by a mass on a bound orbit, the frequencies in the decomposition 
above will be an infinite discrete set and the inverse Fourier transform integrals in Eq.~\eqref{eq:psiDef} and Eq.~\eqref{eq:TDef} are replaced with sums.  In the case of inclined spherical orbits, the discrete spectrum only depends on azimuthal number $m$ and an integer $k$ that gives harmonics of the polar motion,
\begin{equation}
    \omega_{mk} = m\Omega_\phip+k\Omega_\theta .
\end{equation}
Because the integral over $\omega$ is replaced by a sum over $k$, we often make a slight change in notation when restricted to the spherical case, replacing subscripts of the form
$\ell m \omega$ with $\ell mk$, with notational modifications  e.g. ${}_{-2}\psi_{\ell m\omega} (r) \to {}_{-2}\psi_{\ell mk}(r)$, ${}_{-2}T_{\ell m\omega} (r) \to {}_{-2}T_{\ell mk}(r)$, and ${}_{-2}S_{\ell m}(\theta, \varphi; a\omega) \to {}_{-2}S_{\ell mk}(\theta,\varphi)$.

\subsection{Solutions of the Teukolsky equation with a point-particle source}

Inhomogeneous solutions of the Teukolsky equation may be obtained using a Green function,
\begin{align}
    \Psi_4(x) &= 8 \pi \int G_4 (\cS_4 T) \sqrt{-g} \, d^4 x' \nonumber \\
    &=  8 \pi \int (\cS_4^\dag G_4)^{\mu \nu} T_{\mu \nu} \sqrt{-g} \, d^4 x',
\end{align}
where $G_4(x,x')$ is the retarded Green function for the Teukolsky equation and $\cS_4^\dag$ is the adjoint of $\cS_4$. Decomposing this into modes and using the fact that the mode-decomposed Green function can be written in factorised form, we arrive at an expression for the inhomogeneous mode solutions of the form
\begin{align}
    {}_s \psi_{\ell m\omega}(r) = & {}_s C_{\ell m\omega}^{\mathrm{up}}(r) {}_s R_{\ell m\omega}^{\mathrm{up}}(r)
    + {}_s C_{\ell m\omega}^{\mathrm{in}}(r) {}_s R_{\ell m\omega}^{\mathrm{in}}(r),
\end{align}
where ${}_s R_{\ell m\omega}^{\mathrm{up}}(r)$ is a homogeneous solution to Eq.~\eqref{eq:radTeukEq} representing an outgoing wave at infinity and ${}_s R_{\ell m\omega}^{\mathrm{in}}(r)$ is a homogeneous solution representing an ingoing wave at the horizon.

The stress-energy of a point particle on a geodesic with four-velocity $u^\mu$ and position $x_p^\mu = (t_p, r_p, \theta_p, \varphi_p)$ is
\begin{equation}
    T_{\mu \nu} = \frac{\mu}{M} \frac{u_\mu u_\nu}{\Sigma u^t} \delta(r-r_p) \delta(\theta-\theta_p)  \delta(\varphi-\varphi_p).
\end{equation}
For bound motion, this source has non-zero support only within a libration region defined by $\rmin \le \rp \le \rmax$ and $\thetamin \le \thetap \le \thetamax$. Outside this region $T_{\mu \nu} = 0$ and the weighting functions ${}_s C_{\ell m\omega}^{{\mathrm{in}}/{\mathrm{up}}}(r)$ are equal to the constant asymptotic amplitudes ${}_s Z_{\ell m\omega}^{\cH/\infty}$.

In the spherical orbit case the Green function integral is over a sphere of constant radius $\rp = pM$. At the level of modes we then obtain the asymptotic amplitudes as an integral over $\chi$ of a mode-expanded version of $(\cS_4^\dag G_4)^{\mu \nu} T_{\mu \nu}$. The integrand is an explicit function of the orbital parameters $\{\cE, \cLz, \cQ\}$ and $\{\tp(\chi), \rp = pM, \thetap(\chi), \phip(\chi)\}$, of the frequencies $\{\Upsilon_\theta, \Upsilon_\phip, \Upsilon_t\}$, of the spin-weighted spheroidal harmonic and its first and second derivatives evaluated at $\{\theta(\chi), \phip(\chi)\}$, and of the radial Teukolsky function ($R_{\ell mk}^\mathrm{in}$ for $Z_{\ell mk}^{\infty}$ and $R_{\ell mk}^\mathrm{up}$ for $Z_{\ell mk}^{\cH}$) and its first and second derivatives evaluated at $\rp = pM$ \cite{SasaTago03}.

With the asymptotic amplitudes in hand, the fluxes of energy and angular momentum can then be computed using \cite{TeukPres74}
\begin{align}
    \left\langle\frac{dE}{dt}\right\rangle_{\infty} &= \sum_{\ell mk} \frac{1}{4\pi \omega_{mk}^2}\left|Z_{\ell mk}^\infty\right|^2, \\
    \left\langle\frac{dE}{dt}\right\rangle_{\mathcal{H}} &= \sum_{\ell mk} \frac{\alpha_{\ell m k}}{4\pi \omega_{mk}^2}\left|Z_{\ell mk}^{\cH}\right|^2,
\end{align}
and 
\begin{align}
    \left\langle\frac{dL_z}{dt}\right\rangle_{\infty} &= \sum_{\ell mk} \frac{m}{4\pi \omega_{mk}^3}\left|Z_{\ell mk}^\infty\right|^2, \\
    \left\langle\frac{dL_z}{dt}\right\rangle_{\mathcal{H}} &= \sum_{\ell mk} \frac{\alpha_{\ell m k} m}{4\pi \omega_{mk}^3}\left|Z_{\ell mk}^{\cH}\right|^2.
\end{align}
where $\alpha_{\ell m k}$ is a constant that depends on $a$, $m$, $\omega$ and ${}_{-2} \lambda_{\ell m}$ \cite{DrasHugh06}.

\subsection{Post-Newtonian expansions}

Our goal is to obtain PN expanded expressions for the fluxes, which in turn requires an expansion of the asymptotic amplitudes $Z_{\ell mk}^{\infty}$ and $Z_{\ell mk}^{\cH}$. We already have PN expansion for the orbital parameters and frequencies. We will now obtain expansions for the solutions to the Teukolsky radial and spin-weighted spheroidal harmonic equations. 

\subsubsection{Teukolsky radial function}

Solutions to the radial Teukolsky equation, Eq.~\eqref{eq:radTeukEq}, can be found using the methods pioneered by Mano, Suzuki, and Takasugi (MST) \cite{ManoSuzuTaka96a,ManoSuzuTaka96b}. The MST method represents the homogeneous solutions as a convergent sum of special functions. There is some freedom in the particular choice of special functions. For the purpose of producing PN expansions we find it convenient to work with Coulomb wave functions $R_{\mathrm{C}}^\nu$ as given in Eq.~(162) of Ref.~\cite{SasaTago03}. Then, the ``in'' solution is\footnote{Note that this normalization differs from Eq.~(166) of \cite{SasaTago03} by a factor of $K_\nu$. It also does not produce a solution with unit transmission coefficient. We choose this particular normalization since the ratio $K_{-\nu-1}/K_\nu$ (also known as ``tidal response function", see e.g. \cite{Bautista:2023sdf}) is significantly easier to compute than $K_\nu$ on its own. We correct for this difference in normalization when computing the fluxes.}
\begin{align}
    &{}_sR^{\mathrm{in}}_{\ell m \omega} = R_\mathrm{C}^{\nu} + \frac{K_{-\nu-1}}{K_\nu} \, R_\mathrm{C}^{-\nu-1}, \label{eq:MSTin}
\end{align}
where $K_\nu$
depends on the frequency but not on the radial variable. Its explicit form is quite lengthy and can be found in Eq.~(165) of \cite{SasaTago03}.

For the ``up'' solution we turn to Eq.~(B.7) of \cite{Thro10}, which we write here in a form that highlights the leading PN structure of $R_\mathrm{C}^{-\nu-1}$:
\begin{align}
    {}_sR^{\mathrm{up}}_{\ell m \omega}&=e^{-\pi  \epsilon - i  \pi s}\frac{\sin{(\pi  (\nu + s-i\epsilon))}}{i\sin{(2  \pi  \nu)}} \times \nonumber \\
    & \left[ R_{\mathrm{C}}^{-\nu-1} + i e^{-i\pi\nu}\frac{\sin{(\pi(\nu-s+i\epsilon))}}{ \sin{(\pi  (\nu + s-i\epsilon))}} R_{\mathrm{C}}^{\nu} \right].\label{eq:MSTup}
\end{align}
Here $\epsilon = 2M\omega$, $\epsilon_+ = (\epsilon+\tau)/2$ and $\nu = \ell + \mathcal{O}(\epsilon^2)$ is the renormalized angular momentum.

The PN expansion of these MST expressions is straightforward, as $r \sim p$ and $\omega \sim \omega_{mk}\sim p^{-3/2}$. Thus the expansions of the homogenous functions can then be constructed order-by-order in $p^{-1}$. We omit the full details of the PN expansion of the radial functions, instead we refer to the reader the Refs.~\cite{KavaOtteWard15,KavaOtteWard16} where the procedure is described more comprehensively. We implemented this procedure using two independently written codes. Explicit expressions for the PN expanded homogeneous solutions described here can generated with the publicly available \texttt{SFPN} package \cite{SFPN}.

\subsubsection{Angular function}
We next address the calculation of the PN expansion of the spin-weighted spheroidal harmonic.

First, we note that the azimuthal dependence factors out: ${}_{s}S_{\ell m}(\theta, \varphi; a\omega) = {}_{s}S_{\ell m}(\theta, 0; a\omega) e^{i m \varphi}$. We thus start by considering the PN expansion of this dependence on the azimuthal coordinate function $\phip(\chi)$. For the purposes of producing expressions that are exact functions of $x$, it turns out to be convenient to write the expansion in the form \cite{GanzETC07,SagoFuji15,FujiShib20,IsoyETC22}
\begin{align}
    e^{i m\phip} &= \left[\frac{x\cos\chi \pm i\sin\chi}{\sin\thetap}\right]^{|m|}  \left[1+\frac{2ia\chi}{p^{3/2}}+\cO\left(\frac{1}{p^{2}}\right)\right], \label{eq:phaseexp}
\end{align} 
where the sign in the first term is determined by the sign of $m$.
We make two observations: (i) this expression is exact in $x$; (ii) the $(\sin\thetap)^{-|m|}$ factor will later cancel against another corresponding factor.

Next, we write the spin-weighted spheroidal harmonics as a series in terms of spin-weighted \textit{spherical} harmonics ${}_{s}Y_{lm}(\theta, \varphi)$:
\begin{align}
    {}_{s}S_{\ell m}(\theta, \varphi; a \omega) = \sum_{j=|s|}^{\infty}d_{j}(a\omega) \ {}_{s}Y_{jm}(\theta, \varphi), \label{eq:swshseries}
\end{align}
where the coefficients $d_{j}(a\omega)$ depend on $s$, $\ell$ and $m$ and can be expanded as a series in $a\omega$, with leading order behavior $d_{j}(a\omega) \sim (a\omega)^{|j-\ell|}$ for a given value of $\ell$. We can therefore rewrite the expansion in the alternative form
\begin{align}
    {}_s S_{\ell m}(\theta, \varphi; a \omega) = \sum_{n=0}^{\infty}(a\omega)^{n}\left[\sum_{j=-n}^{n}\hat{d}_{jn}\, {}_s Y_{\ell+j,m}(\theta, \varphi)\right].
    \label{eq:swshseries2}
\end{align} 
Explicit expressions for the coefficients $\hat{d}_{jn}$ are readily obtained using the \texttt{SpinWeightedSpheroidalHarmonics} package of the Black Hole Perturbation Toolkit \cite{SpinWeightedSpheroidalHarmonics,BHPTK18}. The important feature is that each additional power of $a \omega$ (corresponding to 1.5PN orders) increases the number of spin-weighted spherical harmonics in the sum by 2, one each at the upper and lower bound.

Now, the explicit dependence on $\theta$ in the spin-weighted spherical harmonics is of the form
\begin{equation}
\label{eq:Ylm-structure}
    {}_s Y_{\ell m} (\theta, 0) \propto (\sin\theta)^{||m|-|s||}\sum_{n=0}^{\ell-|m|}c_{n}\, (\cos\theta)^n
\end{equation}
where the $c_n$ are constants. By virtue of Eq.~\eqref{eq:swshseries} a similar structure is also inherited by the spin-weighted spheroidal harmonics. Combining everything, and using Eq.~\eqref{eq:thfunc} to replace $\cos \theta$ with $\cos \chi$, we thus have that the PN-expanded spin-weighted spheroidal harmonics are of the form
\begin{align}
    {}_s S_{\ell m} &(\thetap(\chi), \phip(\chi);a\omega)= \frac{(x\cos\chi \pm i\sin\chi)^{|m|} }{(\sin\theta)^{|m|-||m|-|s||}}\times \notag \\
    &\left(\text{polynomial in } \chi\right) \times(\text{polynomial in } \cos \chi), \label{eq:spheroidalstructure}
\end{align}
where the degree of the polynomial in $\chi$ is determined by the PN order of the expansion of the phase, Eq.~\eqref{eq:phaseexp}, and the degree of the polynomial in $\cos \chi$ is determined by $\ell$, $m$ and the PN order (via powers of $a\omega$).
\\
\subsection{Sum over modes}
\label{sec:mode-sum}
The mode-sum in Eq.~\eqref{eq:psiDef} is formally a double infinite sum over $\ell$ and $k$. In practice, however, to a given PN order both sums are finite. It is straightforward to see that this is the case for the sum over $\ell$, as $\ell$ determines the leading order PN behavior of the Teukolsky radial functions \cite{KavaOtteWard15,KavaOtteWard16,Munn20}.

A similar truncation rule also holds for the set $\mathbb{K}$ of $k$ modes that must be included at a given PN order. To see this, we start with the structure of the PN-expanded spheroidal harmonics given in Eq.~\eqref{eq:spheroidalstructure}.
The actual integral of interest for computing the asymptotic amplitudes includes these spin-weighted spheroidal harmonics in the integrand along with additional functional dependence on $\chi$ that arises from the presence of a $\theta$-dependent differential operator and a $\theta$-dependent stress-energy tensor. The net result is an integral which can be manipulated into the form 
\begin{align}
    Z_{\ell mk} &= \int_0^{2\pi} \sum_{n} c_{\ell mk}^{(n)} e^{in\chi} \, d\chi  \nonumber \\
    &=2\pi c_{\ell mk}^{(0)}.
\end{align}
Crucially, to a given PN order denoted by $N$, $c_{\ell mk}^{(0)}$ is only required for a specific set of $k$ modes:
\begin{align}
    \mathbb{K} = \{k \in \mathbb{Z} :-\ell-m-2\lfloor N/4\rfloor \leq k \leq \ell-m+2\lfloor N/4\rfloor\}.
\end{align}
The other modes are either identically zero or can be determined from the identity
\begin{equation}
Z_{\ell -m -k} = (-1)^{\ell+k} \bar{Z}_{\ell m k}.
\end{equation}

\begin{widetext}

\section{Results}

\subsection{Structure}

The PN structures of the gravitational fluxes have known general forms \cite{Blan14}.  Based on this expectation, the infinity-side gravitational energy flux can be expressed as
\begin{align}
    \begin{autobreak}
    \left\langle \frac{dE}{dt} \right\rangle_{\infty} =
    \frac{32}{5} \left(\frac{\mu}{M}\right)^2 p^{-5}\bigg[\mathcal{A}_0 
    + \mathcal{A}_1 p^{-1} 
    + \mathcal{A}_{3/2} p^{-3/2} 
    + \mathcal{A}_{2} p^{-2} 
    + \mathcal{A}_{5/2}p^{-5/2} 
    + \left(\mathcal{A}_{3} + \mathcal{A}_{3L}\log(p)\right)p^{-3}
    + \mathcal{A}_{7/2}p^{-7/2}
    + \left(\mathcal{A}_{4}+\mathcal{A}_{4L}\log(p)\right)p^{-4}
    + \left(\mathcal{A}_{9/2}+\mathcal{A}_{9/2L}\log(p)\right)p^{-9/2}
    + \left(\mathcal{A}_{5}+\mathcal{A}_{5L}\log(p)\right)p^{-5}
    + \left(\mathcal{A}_{11/2}+\mathcal{A}_{11/2L}\log(p)\right)p^{-11/2}
    + \left(\mathcal{A}_{6}+\mathcal{A}_{6L}\log(p) + \mathcal{A}_{6L2}\log^{2}(p)\right)p^{-6} 
    + \cdots\bigg],
    \end{autobreak}
\end{align}
with each coefficient $\mathcal{A}_{mLn}$ representing a function of spin $a$ and inclination $x$.  The tag $m$ represents the integer, or half-integer, relative PN order of the term (with $m$ not to be confused with its previous use as azimuthal number).  A script of the form $Ln$ also appears if a power-of-log term, $\log^n(p)$, is present. For simplicity, we abbreviate $\mathcal{A}_{mL0}=\mathcal{A}_{m}$ and $\mathcal{A}_{mL1}=\mathcal{A}_{mL}$.   The non-spinning parts of the gravitational flux for a circular orbit have been previously calculated up to 
22 PN orders \cite{Fuji12b}. 

The general expression for the angular momentum flux at infinity has a similar form 
\begin{align}
    \begin{autobreak}
        \left\langle \frac{dL_z}{dt} \right\rangle_{\infty} = 
            \frac{32}{5}\frac{\mu^2}{M} x p^{-7/2}\bigg[\mathcal{C}_0
            +\mathcal{C}_1 p^{-1}
            +\mathcal{C}_{3/2} p^{-3/2}
            +\mathcal{C}_{2} p^{-2}
            +\mathcal{C}_{5/2} p^{-5/2}
            +\left(\mathcal{C}_{3}+\mathcal{C}_{3L}\log(p)\right)p^{-3}
            +\mathcal{C}_{7/2}p^{-7/2}
            +\left(\mathcal{C}_{4}+\mathcal{C}_{4L}\log(p)\right)p^{-4}
            +\left(\mathcal{C}_{9/2}+\mathcal{C}_{9/2L}\log(p)\right)p^{-9/2}
            +\left(\mathcal{C}_{5}+\mathcal{C}_{5L}\log(p)\right)p^{-5}
            +\left(\mathcal{C}_{11/2}+\mathcal{C}_{11/2L}\log(p)\right)p^{-11/2}
            +\left(\mathcal{C}_{6}+\mathcal{C}_{6L}\log(p)+\mathcal{C}_{6L2}\log^{2}(p)\right)p^{-6}
            +\cdots\bigg].
    \end{autobreak}
\end{align}
We note that we have pulled out an overall prefactor of $x$ in the angular momentum flux. The orbital angular momentum $\mathcal{L}_z$ is defined relative to the spin axis of rotation, with $x = 1$ corresponding to a circular prograde equatorial orbit.  Many, but not all, of the angular momentum flux terms are modulated by the degree of inclination of the orbit, as detailed below. 
\end{widetext}

In the sections that follow, we present the first part of the gravitational energy and angular momentum flux expressions up to 3.5PN in the text of the paper, then provide some number of additional, more complicated terms in tables, and finally discuss in the text a few additional higher-order terms of note.  We limit the presentation of higher-order terms in the paper, as the flux expressions become increasingly unwieldy with increasing PN order. However, the full flux expressions calculated to 12PN order can be found in online repositories \cite{BHPTK18,PostNewtonianSelfForce,UNCGrav22}.

To facilitate this examination, the flux components $\mathcal{A}_{mLn}$ and $\mathcal{C}_{mLn}$ are further broken down into 
\begin{align}
    \mathcal{A}_{mLn}(a,x) &= \mathcal{A}_{mLn}^{(0)} + \sum_{k=0}\mathcal{A}_{mLn}^{Sk}(a,x), \label{eq:ECBD} \\
    \mathcal{C}_{mLn}(a,x) &= \mathcal{C}_{mLn}^{(0)} + \sum_{k=0}\mathcal{C}_{mLn}^{Sk}(a,x), \label{eq:AMCBD}
\end{align}
where $\mathcal{A}_{mLn}^{(0)}$ and $\mathcal{C}_{mLn}^{(0)}$ are the Schwarzschild limit of $\mathcal{A}_{mLn}$ and $\mathcal{C}_{mLn}$ respectively, while $\mathcal{A}_{mLn}^{Sk}$ and $\mathcal{C}_{mLn}^{Sk}$ are components proportional to powers $a^k$ of the spin, making clear terms that vanish in the $a \rightarrow 0$ limit. Finally, to simplify the presentation, we introduce the dimensionless spin parameter $\tilde{a} = a/M$.

\subsection{Gravitational energy flux at infinity}

Through the first 3.5 PN orders, we find the gravitational energy flux for inclined spherical orbits to be
\begin{align}
    \begin{autobreak}
    \MoveEqLeft
    \left\langle \frac{dE}{dt} \right\rangle_{\infty} =
        \frac{32}{5}\frac{\mu^2}{M^2} p^{-5} \bigg[ 1
            - \frac{1247}{336} p^{-1}
            + (4\pi 
            -\frac{73}{12}\tilde{a}x)p^{-3/2}
            + \bigg( -\frac{44711}{9072}
            -\frac{329\tilde{a}^2}{96}
            +\frac{527\tilde{a}^2x^2}{96}\bigg)p^{-2}
            + \bigg(-\frac{8191\pi}{672}
            +\frac{3749}{336}\tilde{a}x\bigg)p^{-5/2}
            + \bigg(\frac{6643739519}{69854400} 
            +\frac{135}{8}\tilde{a}^2
            -\frac{1712}{105}\gamma
            +\frac{856}{105}\log(p)
            +\frac{16}{3}\pi^2
            -\frac{169}{6}\tilde{a}\pi x
            +\frac{73}{21}\tilde{a}^2x^2
            -\frac{3424}{105}\log(2)\bigg)p^{-3}
            +\bigg(-\frac{16285\pi}{504}
            -\frac{809\tilde{a}^2\pi}{48}
            +\frac{83819\tilde{a}x}{1296}
            +\frac{1195}{48}\tilde{a}^3x
            +\frac{1199}{48}\tilde{a}^2\pi x^2
            -\frac{1799}{48}\tilde{a}^3x^3\bigg)p^{-7/2}
        +O(p^{-4})\bigg].
    \end{autobreak} \label{eq:energyfluxI}
\end{align}
We confirm that the terms that survive in the $a=0$ limit match known results \cite{Fuji12b} all the way to 
12 PN order.  Likewise, if we retain nonvanishing $a$ but set $x = 1$ for equatorial circular motion, our results are a complete match to those of \cite{Fuji15}.  For more general, inclined orbits, we find multiple terms proportional to $\tilde{a} x$ or 
odd powers thereof.  These terms reflect ones that switch sign between prograde $0 < x \le 1$ and retrograde $-1 \le x < 0$ orbits (when $a$ is held positive).  We also find that at each order in the expansion $x$ enters as a finite polynomial, reflecting a simple, exact functional dependence on inclination.

\setlength\extrarowheight{3pt}
\begin{table*}[t]
    \centering
    \caption{List of higher-order components of the gravitational energy flux at infinity from 4PN up to 5 PN and a select 6.5PN flux component.}
    \begin{tabular}{|c||c|}
        \hline 
        Component & Energy Flux Expression \\[3pt]
        \hline
        $\mathcal{A}_{4}^{(0)}$ & $-\frac{323105549467}{3178375200}+\frac{232597 \gamma }{4410}-\frac{1369 \pi
   ^2}{126}+\frac{39931 \log (2)}{294}-\frac{47385 \log (3)}{1568}$ \\[3pt] 
        \hline
        $\mathcal{A}_{4L1}^{(0)}$ & $-\frac{232597}{8820}$ \\[3pt]
        \hline
        $\mathcal{A}_{4}^{S1}$ & $\frac{3389}{96}\tilde{a}\pi x$ \\[3pt]
        \hline
        $\mathcal{A}_{4}^{S2}$ & -$\tilde{a}^2\left(\frac{374093}{18144}-\frac{125911}{18144}x^2\right)$ \\[3pt]
        \hline
        $\mathcal{A}_{4}^{S4}$ & $\tilde{a}^4\left(\frac{10703}{768}-\frac{13595 x^2}{384}+\frac{17303 x^4}{768}\right)$ \\[3pt] \hline 
        $\mathcal{A}_{9/2}^{(0)}$ & $\frac{265978667519 \pi }{745113600}-\frac{6848 \gamma  \pi }{105}-\frac{13696}{105} \pi 
   \log (2)$ \\[3pt] \hline 
        $\mathcal{A}_{9/2L1}^{(0)}$ & $\frac{3424 \pi }{105}$ \\[3pt]
        \hline
        $\mathcal{A}_{9/2}^{S1}$ & $-\frac{343985009 \tilde{a} x}{498960}+\frac{1369 \tilde{a} \gamma  x}{9}-\frac{385}{9} \tilde{a} \pi ^2
   x+\frac{95723}{315} \tilde{a} x \log (2)$ \\[3pt] \hline
        $\mathcal{A}_{9/2L1}^{S1}$ & $-\frac{1369}{18} \tilde{a}x$ \\[3pt] \hline
        $\mathcal{A}_{9/2}^{S2}$ & $\frac{216403 \tilde{a}^2 \pi }{2688}+\frac{96937 \tilde{a}^2 \pi  x^2}{2688}$ \\[3pt] \hline
        $\mathcal{A}_{9/2}^{S3}$ & $-\frac{2057}{96}  \tilde{a}^3 x-\frac{12175}{224}\tilde{a}^3x^3$ \\[3pt] \hline
        $\mathcal{A}_{5}^{(0)}$ & $-\frac{2500861660823683}{2831932303200}+\frac{916628467 \gamma }{7858620}-\frac{424223
   \pi ^2}{6804}-\frac{83217611 \log (2)}{1122660}+\frac{47385 \log (3)}{196}$ \\[3pt]
    \hline
    $\mathcal{A}_{5L1}^{(0)}$ & $-\frac{916628467}{15717240}$ \\[3pt]
    \hline
    $\mathcal{A}_{5}^{S1}$ & $\frac{1049395}{3024}\tilde{a}\pi x$ \\[3pt]
    \hline
    $\mathcal{A}_{5}^{S2}$ & $\tilde{a}^2\left(-\frac{62206109341}{139708800}+\frac{204691 \gamma }{2520}-\frac{1913 \pi
   ^2}{72}+\frac{3712887509 x^2}{12700800}-\frac{287509 \gamma  x^2}{2520}+\frac{2687 \pi
   ^2 x^2}{72}+\frac{410131 \log (2)}{2520}-\frac{115025}{504} x^2 \log (2)\right)$ \\[3pt]
    \hline
    $\mathcal{A}_{5L1}^{S2}$ & $\tilde{a}^2\left(-\frac{204691}{5040}+\frac{287509 x^2}{5040}\right)$ \\[3pt]
    \hline
    $\mathcal{A}_{5}^{S3}$ & $\tilde{a}^3\pi\left(\frac{5549 \pi  x}{48}-\frac{8299 \pi  x^3}{48}\right)$, \\[3pt]
    \hline
    $\mathcal{A}_{5}^{S4}$ & $\tilde{a}^4\left(-\frac{256201}{2688}+\frac{85507 x^2}{1344}+\frac{73049 x^4}{896}\right)$ \\[3pt] 
    \hline
    $\mathcal{A}_{13/2}^{S1}$ & $-\frac{1258752377510003 x}{157329572400}-\frac{2695926721 \gamma 
   x}{1746360}+\frac{1352455 \pi ^2 x}{2268}+\frac{257721407 x \log
   (2)}{1746360}-\frac{4208517 x \log (3)}{1760}+\frac{256}{15} x \log (\kappa )$ \\
   & $+ \frac{64}{15}(x-x^3)\Psi^{(0,1)}(\tilde{a})+\frac{64}{15}(x+x^3)\Psi^{(0,2)}(\tilde{a})$ \\[3pt] 
    \hline
    \end{tabular}
    \label{tab:EFluxComp}
\end{table*}

Higher-order components of the energy flux up through 5PN (along with one 6.5PN term) are presented in Table \ref{tab:EFluxComp}. Beyond the complete list to 5PN order, we present and discuss only a few additional 
notable terms, with the remaining terms through 12PN relative order relegated to online repositories.  At higher PN order we begin to find the appearance of non-polynomial functions of $a$ in the form of combinations of polygamma functions, $\psi^{(q)}(z)$ \cite{DLMF}, where $q$ is the integer order.  Only certain even and odd (real) combinations of the polygamma functions appear, which we denote by
\begin{align}
    \Psi^{(q,\sigma)}(\tilde{a}) &= \psi^{(q)}\left(1+\frac{i\sigma\tilde{a}}{\kappa}\right) + \psi^{(q)}\left(1-\frac{i\sigma\tilde{a}}{\kappa}\right),  \notag \\
    i\bar{\Psi}^{(q,\sigma)}(\tilde{a}) &= \psi^{(q)}\left(1+\frac{i\sigma\tilde{a}}{\kappa}\right)-\psi^{(q)}\left(1-\frac{i\sigma\tilde{a}}{\kappa}\right), \label{eq:polygamma}
\end{align}
where the complex argument depends on $\tilde{a}$ and another integer $\sigma$.  The polygamma functions arise from Taylor expansions of the gamma function in the complex plane when $\tilde{a}$ is nonvanishing.
Starting at $\mathcal{A}_{13/2}^{S1}$ (given in the table) and onward, these polygamma functions become increasingly common.  The presence of non-polynomial functions of $a$ in the higher-order components of the gravitational flux has been seen before in both dissipative \cite{TaraETC13,Shah14,Fuji15} and conservative \cite{KavaOtteWard16} quantities. 

\subsection{Gravitational angular momentum flux at infinity}

For the angular momentum flux, we present initially the first 3.5 PN orders
\begingroup
\allowdisplaybreaks
\begin{align}
    \begin{autobreak}
        \MoveEqLeft
        \left\langle\frac{dL_z}{dt}\right\rangle_\infty = \frac{32}{5}\frac{\mu^2}{M} x p^{-7/2} \bigg[ 
            1 - \frac{1247}{336}p^{-1}
            +\bigg(4\pi
            +\frac{61\tilde{a}}{24x}
            -\frac{61}{8}\tilde{a}x\bigg)p^{-3/2}
            +\left(-\frac{44711}{9072}-\frac{65}{16}\tilde{a}^2+\frac{49}{8}\tilde{a}^2x^2\right)p^{-2}
            +\bigg( -\frac{8191\pi}{672}
            -\frac{2633\tilde{a}}{224x}
            +\frac{4301}{224}\tilde{a}x\bigg)p^{-5/2}
            +\bigg(\frac{6643739519}{69854400}
            +\frac{1565\tilde{a}^2}{672}
            -\frac{1712}{105}\gamma
            +\frac{856}{105}\log(p)
            +\frac{16\pi^2}{3}
            +\frac{145\tilde{a}\pi}{12x}
            -\frac{145}{4}\tilde{a}\pi x
            +\frac{8023\tilde{a}^2x^2}{672}
            +\frac{3424}{105}\log(2)\bigg)p^{-3}
            +\bigg(-\frac{16285\pi}{504}
            -\frac{167}{8}\tilde{a}^2\pi
            -\frac{72563\tilde{a}}{6048x}
            -\frac{1223\tilde{a}^3}{192x}
            +\frac{144637\tilde{a}x}{2016}
            +\frac{4481\tilde{a}^3x}{96}
            +29\tilde{a}^2\pi x^2
            -\frac{3253\tilde{a}^3x^3}{64}
            \bigg)p^{-7/2}
        +O(p^{-4})\bigg].
    \end{autobreak}
\end{align}
\endgroup
As mentioned previously, the presence of $x$ in the prefactor reflects the fact that many terms in the 
angular momentum flux flip sign for a retrograde orbit, which means they vanish for polar orbits ($x = 0$).  However, as can be seen, there are some terms with a compensating $1/x$ that lead to angular momentum flux even in the case of a polar orbit.  We find that the leading behavior of these polar-orbit $x = 0$ flux 
terms is
\begin{align} 
    \begin{autobreak}
    \MoveEqLeft
    \lim_{x \to 0} \left\langle\frac{d L_z}{dt}\right\rangle_{\infty} = \frac{32}{5} \frac{\mu^2}{M}\tilde{a}p^{-5} \bigg[
        \frac{61}{24} - \frac{2633}{224}p^{-1}
        + \frac{145\pi}{12}p^{-3/2}
        + \left(-\frac{72563}{6048}-\frac{1223}{192}\tilde{a}^2\right)p^{-2}
    +O(p^{-5/2})\bigg].
    \end{autobreak} \label{eq:angfluxPolar}
\end{align}
The polar-orbit angular momentum flux at infinity vanishes completely in the Schwarzschild limit $\tilde{a} \rightarrow 0$.  Thus, these polar orbit terms can be interpreted as being induced by the frame dragging of the primary Kerr black hole. The frame dragging spurs a precession of the orbit (Eq.~\eqref{eq:AzOpol}) about the axis of rotation of the Kerr black hole, which in turn contributes to the angular momentum flux of the orbit, though at relatively higher PN order.   

Higher-order components of the gravitational angular momentum flux at infinity are presented in Table \ref{tab:LFluxComp}, complete through 5PN.  The entire angular momentum flux at infinity through 12PN can be found in the online repositories \cite{BHPTK18,PostNewtonianSelfForce,UNCGrav22}.

\subsection{Leading-spin terms}

We point out one other feature of the gravitational fluxes, which is what might be called the leading-spin terms in the flux. As we move through the PN expansion, the leading-spin terms represent the first appearance of a new power of spin, $a^n$.  The leading-spin terms are relatively simple polynomials in $x$ and it is possible to list all of them up through 8 PN order.  In the energy flux we find

\begin{table*}[t]
    \centering
    \caption{List of higher-order components of the gravitational angular momentum flux from 4PN up to 5PN.}
    \begin{tabular}{| c || c |}
       \hline
        Component & Angular Momentum Flux Expression \\[3pt]
        \hline
        $\mathcal{C}_{4}^{(0)}$ & $-\frac{323105549467}{3178375200}+\frac{232597 \gamma }{4410}-\frac{1369 \pi^2}{126}+\frac{39931 \log (2)}{294}-\frac{47385 \log (3)}{1568}$ \\[3pt]
        \hline
        $\mathcal{C}_{4L}^{(0)}$ & $-\frac{232597}{8820}$ \\[3pt] 
        \hline
        $\mathcal{C}_{4}^{S1}$ & $\tilde{a}\left(-\frac{16481 \pi }{336 x}+\frac{24247 \pi  x}{336}\right)$ \\[3pt] 
        \hline
        $\mathcal{C}_{4}^{S2}$ & $\tilde{a}^2\left(\frac{255545}{4536}-\frac{88993 x^2}{1512}\right)$ \\[3pt] 
        \hline
        $\mathcal{C}_{4}^{S4}$ & $\tilde{a}^4\left(\frac{685}{64}-\frac{1023 x^2}{32}+\frac{1429 x^4}{64}\right)$ \\[3pt] 
        \hline 
        $\mathcal{C}_{9/2}^{(0)}$ & $\frac{265978667519 \pi }{745113600}-\frac{6848 \gamma  \pi }{105}-\frac{13696}{105} \pi \log (2)$ \\[3pt] 
        \hline
        $\mathcal{C}_{9/2L}^{(0)}$ & $\frac{3424 \pi }{105}$ \\[3pt]
        \hline
        $\mathcal{C}_{9/2}^{S1}$ & $\tilde{a}\left(\frac{23093236423}{69854400 x}-\frac{32699 \gamma }{630 x}+\frac{337 \pi ^2}{18
   x}-\frac{1794649949 x}{1940400}+\frac{39419 \gamma  x}{210}-\frac{337 \pi ^2
   x}{6}-\frac{65291 \log (2)}{630 x}+\frac{78731}{210} x \log (2)\right)$ \\[3pt]
        \hline
        $\mathcal{C}_{9/2L}^{S1}$ & $\tilde{a}\left(\frac{32699}{1260 x}-\frac{39419 x}{420}\right)$ \\[3pt]
        \hline
        $\mathcal{C}_{9/2}^{S2}$ & $\tilde{a}^2 \left(-\frac{269 \pi}{56} +\frac{62635 \pi  x^2}{672}\right)$ \\[3pt]
        \hline
        $\mathcal{C}_{9/2}^{S3}$ & $\tilde{a}^3\left(\frac{160465}{5376 x}-\frac{44699 x}{896}-\frac{190255 x^3}{5376}\right)$ \\[3pt]
        \hline
        $\mathcal{C}_{5}^{(0)}$ & $-\frac{2500861660823683}{2831932303200}+\frac{916628467 \gamma }{7858620}-\frac{424223
   \pi ^2}{6804}-\frac{83217611 \log (2)}{1122660}+\frac{47385 \log (3)}{196}$ \\[3pt] 
        \hline
        $\mathcal{C}_{5L}^{(0)}$ & $-\frac{916628467}{15717240}$ \\[3pt]
        \hline
        $\mathcal{C}_{5}^{S1}$ & $\tilde{a}\left(-\frac{189547 \pi }{2016 x}+\frac{2472011 \pi  x}{6048}\right)$ \\[3pt] 
        \hline
        $\mathcal{C}_{5}^{S2}$ & $\tilde{a}^2\bigg(-\frac{13861984201}{34927200}+\frac{43549 \gamma }{420}-\frac{407 \pi
   ^2}{12}+\frac{798505153 x^2}{2587200}-\frac{14338 \gamma  x^2}{105}+\frac{134 \pi ^2
   x^2}{3}+\frac{17441 \log (2)}{84}-\frac{28676}{105} x^2 \log (2)\bigg)$ \\[3pt]
        \hline
        $\mathcal{C}_{5L}^{S2}$ & $\tilde{a}^2\left(-\frac{43549}{840}+\frac{7169 x^2}{105}\right)$ \\[3pt]
        \hline
        $\mathcal{C}_{5}^{S3}$ & $\tilde{a}^3 \left(-\frac{3905 \pi }{96 x}+\frac{12467 \pi  x}{48}-\frac{8583 \pi  x^3}{32}\right)$ \\[3pt] 
        \hline
        $\mathcal{C}_{5}^{S4}$ & $\tilde{a}^4\left(\frac{995}{224}-\frac{104681 x^2}{672}+\frac{63365 x^4}{336}\right)$ \\[3pt]
        \hline
    \end{tabular}
    \label{tab:LFluxComp}
\end{table*}

\begingroup
\allowdisplaybreaks
\begin{align}
    \mathcal{A}_{3/2}^{S1} =& -\frac{73}{12}\tilde{a}x, \\
    \mathcal{A}_{2}^{S2} =& -\tilde{a}^2\left(\frac{329}{96}-\frac{527}{96}x^2\right), \\
    \mathcal{A}_{7/2}^{S3} =& \tilde{a}^3\left(\frac{1195}{48}x - \frac{1799}{48}x^3\right), \\
    \mathcal{A}_{4}^{S4} =& \tilde{a}^4\left(\frac{10703}{768}-\frac{13595 x^2}{384}+\frac{17303 x^4}{768}\right), \\
    \mathcal{A}_{11/2}^{S5} =& \tilde{a}^5\left(-\frac{14011 x}{128}+\frac{54221 x^3}{192}-\frac{68905 x^5}{384}\right), \\
    \mathcal{A}_{6}^{S6} =& \tilde{a}^6\bigg(-\frac{56053}{1536}+\frac{230245 x^2}{1536}-\frac{296219 x^4}{1536} \\
    &+\frac{122027 x^6}{1536}\bigg), \notag \\
    \mathcal{A}_{15/2}^{S7} =& \tilde{a}^7\bigg(\frac{18145 x}{48}-\frac{184973 x^3}{128}+\frac{114529 x^5}{64} \\
    &-\frac{277415 x^7}{384}\bigg), \notag \\
    \mathcal{A}_{8}^{S8} =& \tilde{a}^8\bigg(\frac{7681475}{98304}-\frac{3670673 x^2}{8192}+\frac{14614499 x^4}{16384} \\
    &-\frac{18440531 x^6}{24576}+\frac{7480577 x^8}{32768}\bigg). \notag
\end{align}%
\endgroup
In the angular momentum flux we find
\begingroup
\allowdisplaybreaks
\begin{align}
    \mathcal{C}_{3/2}^{S1} =& \tilde{a}\left(\frac{61}{24 x}-\frac{61 x}{8}\right),\\
    \mathcal{C}_{2}^{S2} =& \tilde{a}^2\left(-\frac{65}{16}+\frac{49 x^2}{8}\right), \\
    \mathcal{C}_{7/2}^{S3} =& \tilde{a}^3\left(-\frac{1223}{192 x}+\frac{4481 x}{96}-\frac{3253 x^3}{64}\right), \\
    \mathcal{C}_{4}^{S4} =& \tilde{a}^4\left(\frac{685}{64}-\frac{1023 x^2}{32}+\frac{1429 x^4}{64}\right), \\
    \mathcal{C}_{11/2}^{S5} =& \tilde{a}^5\bigg(\frac{5795}{384 x}-\frac{61579 x}{384}+\frac{45171 x^3}{128} \\
    &-\frac{81817 x^5}{384}\bigg), \notag \\
    \mathcal{C}_{6}^{S6} =& \tilde{a}^6\bigg(-\frac{5643}{256}+\frac{13163 x^2}{128}-\frac{37403 x^4}{256} \\
    & +\frac{1045 x^6}{16}\bigg), \notag \\
    \mathcal{C}_{15/2}^{S7} =& \tilde{a}^7\bigg(-\frac{667699}{24576 x}+\frac{2647579 x}{6144}-\frac{17566297 x^3}{12288} \notag \\
    &+\frac{10586747 x^5}{6144}-\frac{5712337 x^7}{8192}\bigg),  \\
    \mathcal{C}_{8}^{S8} =& \tilde{a}^8\bigg(\frac{313917}{8192}-\frac{514309 x^2}{2048}+\frac{2238519 x^4}{4096} \\
    &-\frac{1009509 x^6}{2048}+\frac{1304317 x^8}{8192}\bigg) . \notag
\end{align}%
\endgroup
We see that the odd-power in $a$ terms appear in half-PN-order terms, $\mathcal{A}_{2n+3/2}^{S(2n+1)}$ and $\mathcal{C}_{2n+3/2}^{S(2n+1)}$, while the even power in $a$ terms appear in integer-PN-order terms, $\mathcal{A}_{2n}^{S(2n)}$ and $\mathcal{C}_{2n}^{S(2n)}$. The power in $a$ is correlated with the polynomial order in $x$.  This simple polynomial in $x$ structure in the leading-spin terms leads to the question of whether they comprise a set of sequences of functions that might allow the prediction of all leading-spin terms in the gravitational flux.  Such sequences of functions, if they exist, would be akin to leading-logarithm terms and similar sequences present in the gravitational fluxes from eccentric orbits \cite{JohnMcDaShahWhit15,MunnEvan19a,MunnEvan20a}. 

At the moment, extracting sequences directly from such few flux components does not seem possible. To uncover whether there exist analytic sequences for the leading-spin terms in the gravitational fluxes, we may need either a higher-order PN-expansion extrapolation of the results in this paper or a method analogous to \cite{MunnEvan19a,MunnEvan20a}, where low-order full PN theory 
provides information on the source multipole moment dependence of the first terms and hints as to how to extrapolate to all PN orders.

\subsection{Comparison with numerical results}

\begin{figure*}[htb!]
    \includegraphics[width=\textwidth]{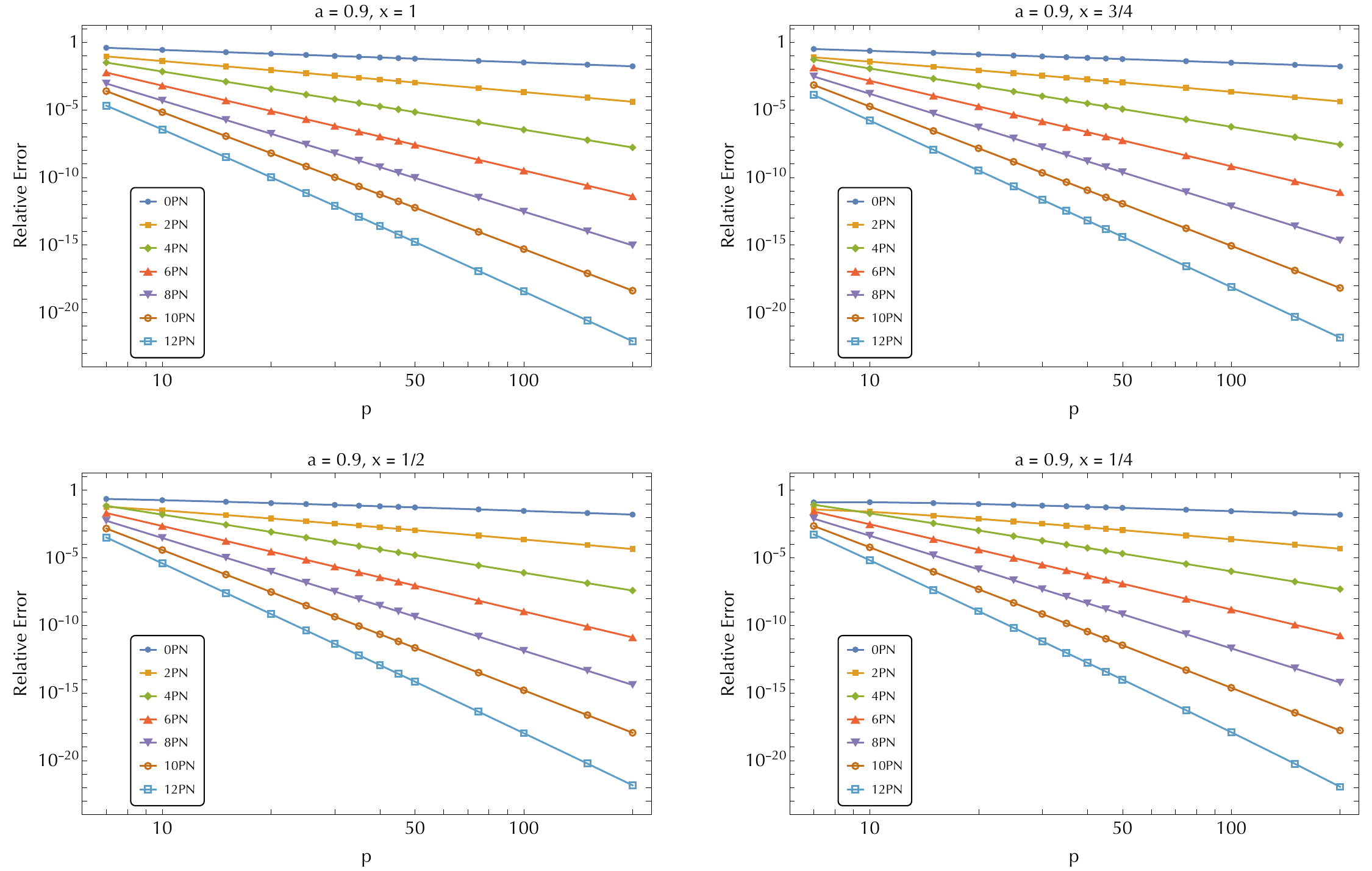}
    \caption{Relative error in the infinity-side energy flux when comparing the numerical evaluation of PN expansions versus numerical fluxes computed with a Teukolsky code. Results are shown as a function of $p$ for orbits with $a=0.9M$ and $x \in \{\frac14, \frac12, \frac34, 1\}$. The different colored points and curves show residuals as expansions of increasing PN order are used, from 0PN to 12PN.  The relative error decreases with larger $p$ as the PN approximation becomes more valid and as more terms are included in the PN expansion.  Comparisons of angular momentum flux at infinity are similar.}
    \label{fig:NumComp}
\end{figure*}

As with all PN expansions, our flux expressions are valid for orbits of sufficiently large separation. They are, however, exact functions of spin $a$ and inclination $x$. To validate the results and probe the applicability of the PN expanded gravitational flux across the orbital parameter space, we compared the PN expansions against numerical flux results obtained using the \texttt{Teukolsky} package of the Black Hole Perturbation Toolkit \cite{TeukolskyBHPT,BHPTK18}. Figure \ref{fig:NumComp} shows a representative comparison in which the energy flux at infinity is compared for orbits with $a=0.9M$ and $x \in \{\frac14, \frac12, \frac34, 1\}$. The fall-off of the relative error becomes more rapid as we include more terms in the expansion, with the residual scaling with $1/p$ at the expected rate for the number of PN terms included in the expansion.

\section{Conclusion and Outlook}

In this work, we presented the gravitational flux from spherical, inclined orbits about a spinning (Kerr) primary black hole, expanded to 12PN. The resulting PN-expanded expression is exact in spin $a$ and inclination parameter $x$.  The fluxes to infinity are discussed in the main body of the paper, while the fluxes into the primary's horizon are more briefly discussed in Appendix \ref{sec:horizon}. A brief discussion on the compatibility with retrograde orbits is given in Appendix \ref{sec:retrograde}. Both complete datasets are provided in online repositories \cite{BHPTK18,PostNewtonianSelfForce,UNCGrav22}.

We highlighted interesting features in the structure of the gravitational fluxes, noting the expected asymmetric dependence of $x$ with  $a$, as well as the presence of a complex relationship between $a$ and $x$ that begins to appear in the higher-order flux components. We also noted the behavior of the leading-spin terms in the gravitational flux.  Whether or not the leading-spin terms comprise a sequence function is the subject of a future study.

The PN-expanded fluxes show excellent agreement with numerically computed values. The rate of fall-off of the residual after subtracting a PN expansion to a given order is as expected, providing convincing confirmation of the correctness of the PN expansions.

Several future directions are now possible. The high-order PN expansion of the fluxes presented here (and the corresponding amplitudes) are likely to be of immediate use in adiabatic waveform models, analogous to \cite{SagoFujiNaka24}. More challenging is the extension of our calculation to the conservative sector, in which regularized quantities local to the worldline are required. There, the two most significant challenges will be: (i) in deriving an appropriate regularization scheme;  and (ii) in obtaining closed-form expressions for sums over $m$ for \textit{generic} $\ell$. Both of these will be addressed in a future work. (Note: As this paper was approaching completion, we became aware of a similar effort. The authors of the other project agreed to a simultaneous submission of papers \cite{SagoFujiNaka24}.)

\section*{Acknowledgements}

We thank Niels Warburton for useful discussions. We also thank Norichika Sago, Ryuichi Fujita, Hiroyuki Nakano and Soichiro Isoyama for helpful correspondence. This research was supported by NSF Grant Nos.~PHY-2110335 and PHY-2409604 to the University of North Carolina at Chapel Hill and the Hamilton Award - University of North Carolina at Chapel Hill. CK and JN acknowledge support from Science Foundation Ireland under Grant number 21/PATH-S/9610. 

\appendix 

\begin{widetext}
    
\section{Horizon Fluxes}
\label{sec:horizon}

For the horizon fluxes, the general PN structure can be written in a manner similar to the infinity-side fluxes.  We expect the energy flux to the horizon to have the form
\begin{align}
    \begin{autobreak}
    \left\langle\frac{dE}{dt}\right\rangle_{\mathcal{H}} = 
    \frac{32}{5}\frac{\mu^2}{M^2}p^{-15/2}\bigg[ 
        \mathcal{B}_0
        +\mathcal{B}_1p^{-1}
        +\mathcal{B}_{3/2}p^{-3/2}
        +\mathcal{B}_{2}p^{-2}
        +\mathcal{B}_{5/2}p^{-5/2}
        +\left(\mathcal{B}_{3} + \mathcal{B}_{3L}\log(p)\right)p^{-3}
        +\mathcal{B}_{7/2}p^{-7/2}
        +\left(\mathcal{B}_{4}+\mathcal{B}_{4L}\log(p)\right)p^{-4}
        +\left(\mathcal{B}_{9/2}+\mathcal{B}_{9/2L}\log(p)\right)p^{-9/2}
        +\left(\mathcal{B}_{5}+\mathcal{B}_{5L}\log(p)\right)p^{-5}
        +\left(\mathcal{B}_{11/2}+\mathcal{B}_{11/2L}\log(p)\right)p^{-11/2}
        +\left(\mathcal{B}_{6}+\mathcal{B}_{6L}\log(p)+\mathcal{B}_{6L2}\log^2(p)\right)p^{-6}
    +\cdots\bigg],
    \end{autobreak}
\end{align} 
and the angular momentum flux to the horizon to be similar, though differing in prefactor
\begin{align}
    \begin{autobreak}
        \left\langle\frac{dL_z}{dt}\right\rangle_{\mathcal{H}} = 
        \frac{32}{5}\frac{\mu^2}{M}xp^{-6}\bigg[
        \mathcal{D}_0
        +\mathcal{D}_1p^{-1}
        +\mathcal{D}_{3/2}p^{-3/2}
        +\mathcal{D}_{2}p^{-2}
        +\mathcal{D}_{5/2}p^{-5/2}
        +\left(\mathcal{D}_{3} + \mathcal{D}_{3L}\log(p)\right)p^{-3}
        +\mathcal{D}_{7/2}p^{-7/2}
        +\left(\mathcal{D}_{4}+\mathcal{D}_{4L}\log(p)\right)p^{-4}
        +\left(\mathcal{D}_{9/2}+\mathcal{D}_{9/2L}\log(p)\right)p^{-9/2}
        +\left(\mathcal{D}_{5}+\mathcal{D}_{5L}\log(p)\right)p^{-5}
        +\left(\mathcal{D}_{11/2}+\mathcal{D}_{11/2L}\log(p)\right)p^{-11/2}
        +\left(\mathcal{D}_{6}+\mathcal{D}_{6L}\log(p)+\mathcal{D}_{6L2}\log^2(p)\right)p^{-6}
        +\cdots\bigg].
    \end{autobreak}
\end{align}

We use a decomposition in the same form as used in Eqs. \eqref{eq:ECBD}-\eqref{eq:AMCBD} to separate the spin-dependent components of $\mathcal{B}_{mLn}$ and $\mathcal{D}_{mLn}$ 
\begin{align}
    \mathcal{B}_{mLn}(\tilde{a},x) = \mathcal{B}_{mLn}^{(0)} + \sum_{k=0}\mathcal{B}_{mLn}^{Sk}(\tilde{a},x), \qquad
    \mathcal{D}_{mLn}(\tilde{a},x) = \mathcal{D}_{mLn}^{(0)} + \sum_{k=0}\mathcal{D}_{mLn}^{Sk}(\tilde{a},x), 
\end{align}
with $\mathcal{B}_{mLn}^{(0)}$ and $\mathcal{D}_{mLn}^{(0)}$ representing the non-spinning limit of $\mathcal{B}_{mLn}$ and $\mathcal{D}_{mLn}$ respectively, while $\mathcal{B}_{mLn}^{Sk}$ and $\mathcal{D}_{mLn}^{Sk}$ are components proportional to $a^k$. 

The horizon flux expressions are far more complex and unwieldy to write down as compared to the infinity-side flux expressions.  We give explicitly here only the leading 1.5PN orders for illustration and point to the online repositories \cite{BHPTK18,PostNewtonianSelfForce,UNCGrav22} for the complete dataset, which is 9.5PN relative to the leading horizon flux or equivalently 12PN relative to the infinity-side flux.  The energy and angular momentum fluxes are
\begingroup
\allowdisplaybreaks
\begin{align}
    \begin{autobreak}
    \MoveEqLeft
    \left\langle\frac{dE}{dt}\right\rangle_{\mathcal{H}} = 
    \frac{32}{5}\frac{\mu^2}{M^2}p^{-15/2}\bigg[ -\frac{1}{4} \tilde{a} x
    -\frac{9 \tilde{a}^3 x}{32}
    -\frac{15 \tilde{a}^3 x^3}{32}
    +\bigg( -\tilde{a} x
    -\frac{81 \tilde{a}^3 x}{32}
    +\frac{15 \tilde{a}^3 x^3}{32} \bigg)xp^{-1}
    +\bigg( \frac{1}{2}
    -\frac{3 \tilde{a}^2}{8}
    -\frac{123 \tilde{a}^4}{64}
    +\frac{179 \tilde{a}^2 x^2}{24}
    +3 \tilde{a}^4 x^2
    +\frac{155 \tilde{a}^4 x^4}{64}
    +\frac{\kappa }{2}
    -\frac{3 \tilde{a}^2 \kappa }{8}
    +\frac{3 \tilde{a}^4 \kappa}{16}
    +\frac{15}{8} \tilde{a}^2 x^2 \kappa 
    -\frac{3}{8} \tilde{a}^4 x^2 \kappa 
    +\frac{3}{16} \tilde{a}^4 x^4 \kappa 
    +\bigg(\frac{\tilde{a}}{4}
    -\frac{3 \tilde{a}^3}{16}
    -\frac{\tilde{a} x^4}{4}
    +\frac{3 \tilde{a}^3 x^4}{16}
    \bigg)\bar{\Psi}^{(0,1)}(\tilde{a}) 
    +\bigg(\frac{\tilde{a}}{8}
    +\frac{3 \tilde{a}^3}{8}
    +\frac{3 \tilde{a} x^2}{4}
    +\frac{9 \tilde{a}^3 x^2}{4}
    +\frac{\tilde{a} x^4}{8}
    +\frac{3 \tilde{a}^3 x^4}{8}\bigg)\bar{\Psi}^{(0,2)}(\tilde{a})
    \bigg)p^{-3/2}
    +O(p^{-2})\bigg],
    \end{autobreak}
\end{align}
\begin{align}
    \begin{autobreak}
    \MoveEqLeft
    \left\langle\frac{dL_z}{dt}\right\rangle_{\mathcal{H}} =
    \frac{32}{5}\frac{\mu^2}{M}p^{-6}\bigg[ -\frac{\tilde{a}}{8}
    -\frac{33 \tilde{a}^3}{128}
    -\frac{\tilde{a} x^2}{8}
    -\frac{9 \tilde{a}^3 x^2}{64}
    -\frac{45 \tilde{a}^3 x^4}{128}
    +\bigg(-\frac{5 \tilde{a}}{4}
    -\frac{375 \tilde{a}^3}{128}
    +\frac{\tilde{a} x^2}{4}
    +\frac{63 \tilde{a}^3 x^2}{64}
    -\frac{15 \tilde{a}^3 x^4}{128}\bigg)p^{-1}
    +\bigg(\frac{x}{2}
    +\frac{63 \tilde{a}^2 x}{16}
    +\frac{15 \tilde{a}^4 x}{16}
    +\frac{139 \tilde{a}^2 x^3}{48}
    +\frac{29 \tilde{a}^4 x^3}{16}
    +\frac{x \kappa }{2}
    +\frac{9}{16} \tilde{a}^2 x \kappa 
    +\frac{15}{16} \tilde{a}^2 x^3 \kappa
    +\bigg(\frac{\tilde{a} x}{4}
    -\frac{3 \tilde{a}^3 x}{16}
    -\frac{\tilde{a} x^3}{4}
    +\frac{3 \tilde{a}^3 x^3}{16}\bigg)\bar{\Psi}^{(0,1)}(\tilde{a}) 
    +\bigg( \frac{\tilde{a} x}{2}
    +\frac{3 \tilde{a}^3 x}{2}
    +\frac{\tilde{a} x^3}{2}
    +\frac{3 \tilde{a}^3 x^3}{2}\bigg)\bar{\Psi}^{(0,2)}(\tilde{a})
    \bigg)p^{-3/2}
    +O(p^{-2})\bigg].
    \end{autobreak}
\end{align}%
\endgroup

Looking at the non-spinning limit, $\tilde{a}\rightarrow 0$, we find the first non vanishing term in the energy flux at $p^{-9}$. As expected this agrees with the fluxes for a particle on a circular orbit in Schwarzschild spacetime \cite{Shah14}.  We confirmed that the Schwarzschild limit of our complete expansions match previous work.

\begin{figure*}[htb!]
    \includegraphics[width=\textwidth]{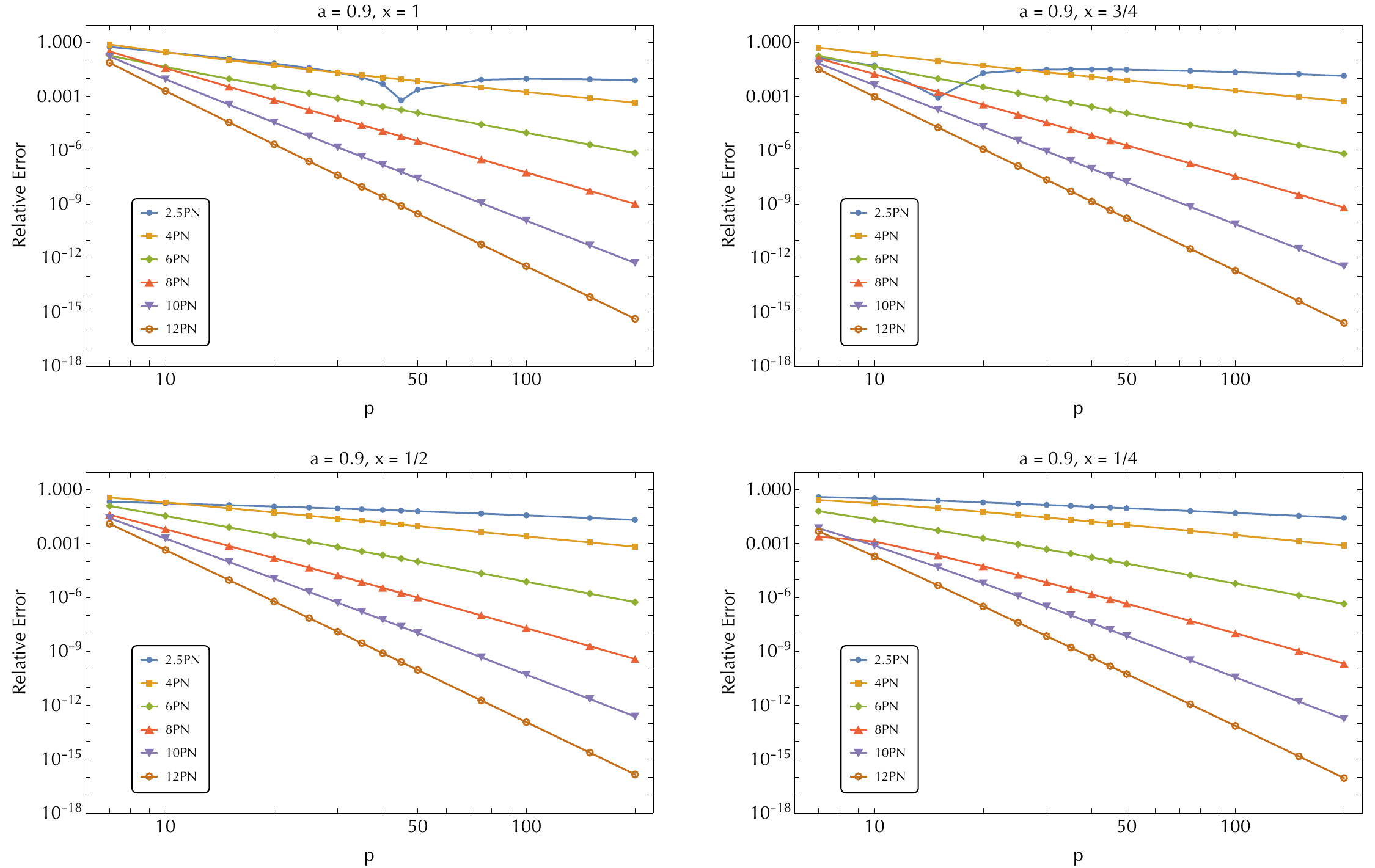}
    \caption{Relative error in the horizon energy flux when comparing the numerical evaluation of PN expansions versus numerical fluxes computed with a Teukolsky code.  Relative error is shown as a function of $p$ for orbits with $a=0.9M$ and $x \in \{\frac14, \frac12, \frac34, 1\}$.  The downward spikes present in the 2.5PN curves that appear in a log-of-absolute-value plot are due to a zero-crossing, $\lim_{p \to p_C}\Delta\dot{E}_{\mathcal{H}} = 0$, at that order of approximation (see text for further discussion). }
    \label{fig:NumCompHor}
\end{figure*}

In Fig.~\ref{fig:NumCompHor}, we present a comparison between the numerically-evaluated PN-expanded horizon fluxes and the full numerical (Teukolsky) horizon fluxes for orbits with $a = 0.9M$ and $x \in \{\frac{1}{4},\frac{1}{2},\frac{3}{4},1\}$. The same general fall-off behavior as seen in the infinity-side fluxes is evident here as we include more terms in the expansion.  One novel feature, not seen in the infinity-side fluxes, is non-uniform behavior of the residuals when using the lowest order, 2.5PN expansion.  As we confirmed, the dip in the residuals corresponds to a zero-crossing in the difference    
\begin{align}
    \lim_{p \to p_C} \left(\left\langle\frac{dE}{dt}\right\rangle_{\mathcal{H}}^{PN} - \left\langle\frac{dE}{dt}\right\rangle_{\mathcal{H}}^{\mathrm{Num}}\right) = 0 
\end{align}
at some critical value $p_C$.  On a log-of-absolute-value plot, a zero-crossing corresponds to a downward spike. The location of $p_C$ depends on $x$.  Smaller values of $|x|$, corresponding to increasingly polar orbits, push $p_C$ towards tighter orbits.  Once several more PN orders are included, the residuals take on more of a power-law appearance.

\section{Retrograde Orbits}
\label{sec:retrograde}

The numerical comparisons displayed in Fig.~\ref{fig:NumComp} and Fig.~\ref{fig:NumCompHor} were for prograde orbits 
($x>0$).  The flux expressions we computed are also compatible with retrograde orbits, and we present them in Fig.~\ref{fig:NumRetroInf} and Fig.~\ref{fig:NumRetroHor}. In general, the large separation behavior for prograde and retrograde orbits remains similar in form and approaching the same limits, and the residuals improve as we include more terms in the expansion. We do note that there are differences in the close separation regime, as the last stable orbits differ markedly between prograde and retrograde orbits, especially for high values of $a$.

\begin{figure*}[htb!]
    \includegraphics[width=\textwidth]{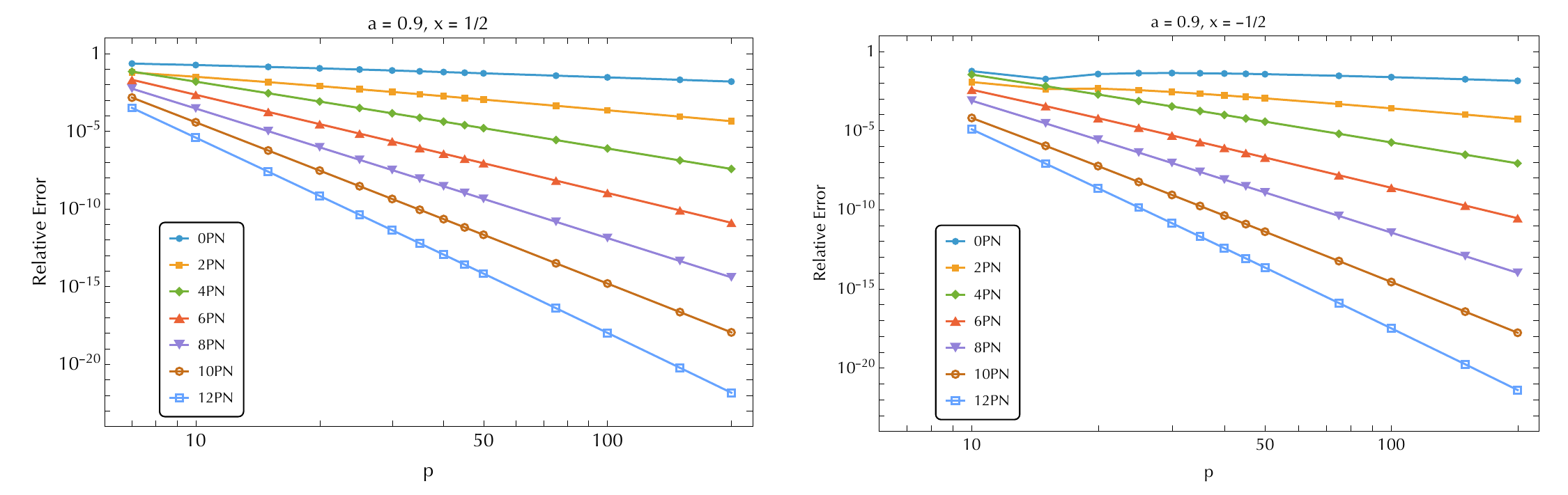}
    \caption{Comparison of relative errors in the infinity-side energy flux for prograde versus retrograde orbits when comparing the numerical evaluation of PN expansions versus numerical fluxes computed with a Teukolsky code. Results are shown as a function of $p$ for orbits with $a=0.9M$ and $x \in \{-\frac12, \frac12\}$. Negative $x$ values represent retrograde orbits. In the retrograde orbit case, we exclude $p = 7$, as the location of the last stable spherical orbit is at $p \backsimeq 7.1029$ \cite{SteiWarb19}.}
    \label{fig:NumRetroInf}
\end{figure*}

\begin{figure*}[htb!]
    \includegraphics[width=\textwidth]{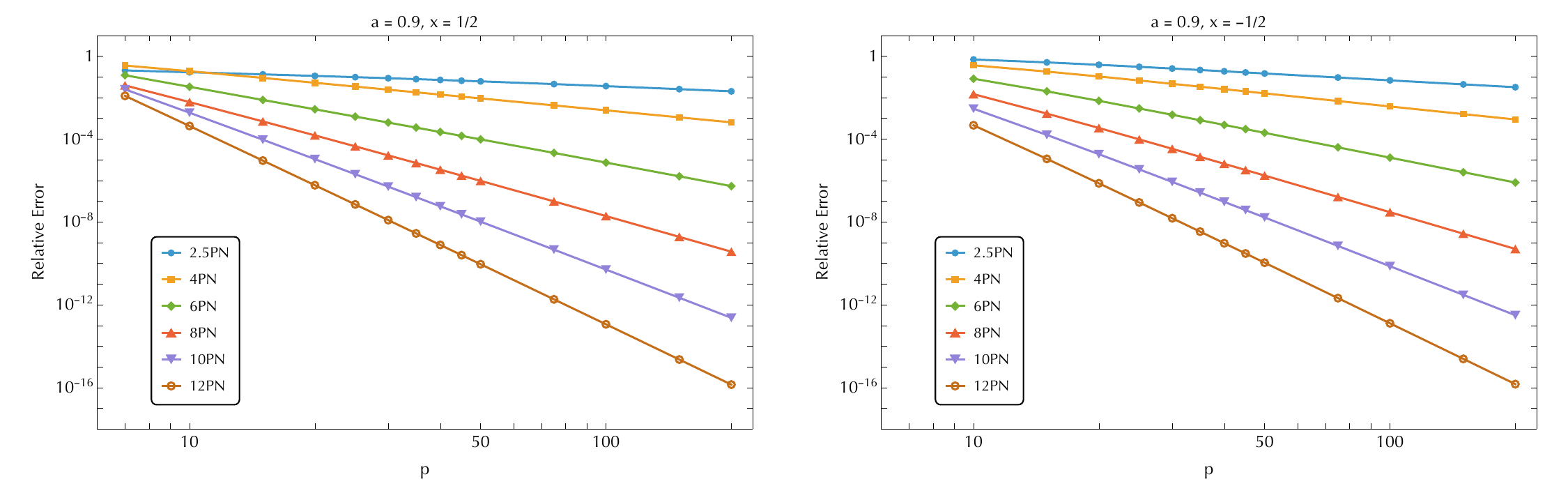}
    \caption{Comparison of relative errors in the horizon energy flux for prograde versus retrograde orbits when comparing the numerical evaluation of PN expansions versus numerical fluxes computed with a Teukolsky code. Results are shown as a function of $p$ for orbits with $a=0.9M$ and $x \in \{-\frac12, \frac12\}$. As with the previous figure, we exclude $p = 7$ in the retrograde orbit plot, as the location of the last stable spherical orbit is at $p \backsimeq 7.1029$ \cite{SteiWarb19}.}
    \label{fig:NumRetroHor}
\end{figure*}
\end{widetext}

\clearpage
\bibliography{references}

\begin{thebibliography}{60}%
\makeatletter
\providecommand \@ifxundefined [1]{%
 \@ifx{#1\undefined}
}%
\providecommand \@ifnum [1]{%
 \ifnum #1\expandafter \@firstoftwo
 \else \expandafter \@secondoftwo
 \fi
}%
\providecommand \@ifx [1]{%
 \ifx #1\expandafter \@firstoftwo
 \else \expandafter \@secondoftwo
 \fi
}%
\providecommand \natexlab [1]{#1}%
\providecommand \enquote  [1]{``#1''}%
\providecommand \bibnamefont  [1]{#1}%
\providecommand \bibfnamefont [1]{#1}%
\providecommand \citenamefont [1]{#1}%
\providecommand \href@noop [0]{\@secondoftwo}%
\providecommand \href [0]{\begingroup \@sanitize@url \@href}%
\providecommand \@href[1]{\@@startlink{#1}\@@href}%
\providecommand \@@href[1]{\endgroup#1\@@endlink}%
\providecommand \@sanitize@url [0]{\catcode `\\12\catcode `\$12\catcode `\&12\catcode `\#12\catcode `\^12\catcode `\_12\catcode `\%12\relax}%
\providecommand \@@startlink[1]{}%
\providecommand \@@endlink[0]{}%
\providecommand \url  [0]{\begingroup\@sanitize@url \@url }%
\providecommand \@url [1]{\endgroup\@href {#1}{\urlprefix }}%
\providecommand \urlprefix  [0]{URL }%
\providecommand \Eprint [0]{\href }%
\providecommand \doibase [0]{http://dx.doi.org/}%
\providecommand \selectlanguage [0]{\@gobble}%
\providecommand \bibinfo  [0]{\@secondoftwo}%
\providecommand \bibfield  [0]{\@secondoftwo}%
\providecommand \translation [1]{[#1]}%
\providecommand \BibitemOpen [0]{}%
\providecommand \bibitemStop [0]{}%
\providecommand \bibitemNoStop [0]{.\EOS\space}%
\providecommand \EOS [0]{\spacefactor3000\relax}%
\providecommand \BibitemShut  [1]{\csname bibitem#1\endcsname}%
\let\auto@bib@innerbib\@empty
\bibitem [{\citenamefont {{Amaro-Seoane}}\ \emph {et~al.}(2007)\citenamefont {{Amaro-Seoane}}, \citenamefont {{Gair}}, \citenamefont {{Freitag}}, \citenamefont {{Miller}}, \citenamefont {{Mandel}}, \citenamefont {{Cutler}},\ and\ \citenamefont {{Babak}}}]{AmarETC07}%
  \BibitemOpen
  \bibfield  {author} {\bibinfo {author} {\bibfnamefont {P.}~\bibnamefont {{Amaro-Seoane}}}, \bibinfo {author} {\bibfnamefont {J.~R.}\ \bibnamefont {{Gair}}}, \bibinfo {author} {\bibfnamefont {M.}~\bibnamefont {{Freitag}}}, \bibinfo {author} {\bibfnamefont {M.~C.}\ \bibnamefont {{Miller}}}, \bibinfo {author} {\bibfnamefont {I.}~\bibnamefont {{Mandel}}}, \bibinfo {author} {\bibfnamefont {C.~J.}\ \bibnamefont {{Cutler}}}, \ and\ \bibinfo {author} {\bibfnamefont {S.}~\bibnamefont {{Babak}}},\ }\href {\doibase 10.1088/0264-9381/24/17/R01} {\bibfield  {journal} {\bibinfo  {journal} {Classical and Quantum Gravity}\ }\textbf {\bibinfo {volume} {24}},\ \bibinfo {pages} {R113} (\bibinfo {year} {2007})},\ \Eprint {http://arxiv.org/abs/astro-ph/0703495} {astro-ph/0703495} \BibitemShut {NoStop}%
\bibitem [{\citenamefont {{{Abbott}, B.~P.~et al.~(The LIGO Scientific Collaboration and VIRGO Collaboration)}}(2016)}]{AbboETC16a}%
  \BibitemOpen
  \bibfield  {author} {\bibinfo {author} {\bibnamefont {{{Abbott}, B.~P.~et al.~(The LIGO Scientific Collaboration and VIRGO Collaboration)}}},\ }\href {\doibase 10.1103/PhysRevLett.116.061102} {\bibfield  {journal} {\bibinfo  {journal} {Physical Review Letters}\ }\textbf {\bibinfo {volume} {116}},\ \bibinfo {eid} {061102} (\bibinfo {year} {2016})},\ \Eprint {http://arxiv.org/abs/1602.03837} {arXiv:1602.03837 [gr-qc]} \BibitemShut {NoStop}%
\bibitem [{\citenamefont {{{Abbott}, B.~P.~et al.}}(2017)}]{AbboETC17}%
  \BibitemOpen
  \bibfield  {author} {\bibinfo {author} {\bibnamefont {{{Abbott}, B.~P.~et al.}}} (\bibinfo {collaboration} {LIGO Scientific Collaboration and Virgo Collaboration}),\ }\href {\doibase 10.1103/PhysRevLett.119.161101} {\bibfield  {journal} {\bibinfo  {journal} {Phys. Rev. Lett.}\ }\textbf {\bibinfo {volume} {119}},\ \bibinfo {pages} {161101} (\bibinfo {year} {2017})}\BibitemShut {NoStop}%
\bibitem [{\citenamefont {Abbott}\ \emph {et~al.}(2021)\citenamefont {Abbott} \emph {et~al.}}]{LIGOScientific:2020ibl}%
  \BibitemOpen
  \bibfield  {author} {\bibinfo {author} {\bibfnamefont {R.}~\bibnamefont {Abbott}} \emph {et~al.} (\bibinfo {collaboration} {LIGO Scientific, Virgo}),\ }\href {\doibase 10.1103/PhysRevX.11.021053} {\bibfield  {journal} {\bibinfo  {journal} {Phys. Rev. X}\ }\textbf {\bibinfo {volume} {11}},\ \bibinfo {pages} {021053} (\bibinfo {year} {2021})},\ \Eprint {http://arxiv.org/abs/2010.14527} {arXiv:2010.14527 [gr-qc]} \BibitemShut {NoStop}%
\bibitem [{\citenamefont {Abbott}\ \emph {et~al.}(2023)\citenamefont {Abbott} \emph {et~al.}}]{KAGRA:2021vkt}%
  \BibitemOpen
  \bibfield  {author} {\bibinfo {author} {\bibfnamefont {R.}~\bibnamefont {Abbott}} \emph {et~al.} (\bibinfo {collaboration} {KAGRA, VIRGO, LIGO Scientific}),\ }\href {\doibase 10.1103/PhysRevX.13.041039} {\bibfield  {journal} {\bibinfo  {journal} {Phys. Rev. X}\ }\textbf {\bibinfo {volume} {13}},\ \bibinfo {pages} {041039} (\bibinfo {year} {2023})},\ \Eprint {http://arxiv.org/abs/2111.03606} {arXiv:2111.03606} \BibitemShut {NoStop}%
\bibitem [{\citenamefont {Gair}\ \emph {et~al.}(2017)\citenamefont {Gair}, \citenamefont {Babak}, \citenamefont {Sesana}, \citenamefont {Amaro-Seoane}, \citenamefont {Barausse}, \citenamefont {Berry}, \citenamefont {Berti},\ and\ \citenamefont {Sopuerta}}]{GairETC17}%
  \BibitemOpen
  \bibfield  {author} {\bibinfo {author} {\bibfnamefont {J.~R.}\ \bibnamefont {Gair}}, \bibinfo {author} {\bibfnamefont {S.}~\bibnamefont {Babak}}, \bibinfo {author} {\bibfnamefont {A.}~\bibnamefont {Sesana}}, \bibinfo {author} {\bibfnamefont {P.}~\bibnamefont {Amaro-Seoane}}, \bibinfo {author} {\bibfnamefont {E.}~\bibnamefont {Barausse}}, \bibinfo {author} {\bibfnamefont {C.~P.~L.}\ \bibnamefont {Berry}}, \bibinfo {author} {\bibfnamefont {E.}~\bibnamefont {Berti}}, \ and\ \bibinfo {author} {\bibfnamefont {C.}~\bibnamefont {Sopuerta}},\ }\href {\doibase 10.1088/1742-6596/840/1/012021} {\bibfield  {journal} {\bibinfo  {journal} {Journal of Physics: Conference Series}\ }\textbf {\bibinfo {volume} {840}},\ \bibinfo {pages} {012021} (\bibinfo {year} {2017})}\BibitemShut {NoStop}%
\bibitem [{\citenamefont {Barack}(2009)}]{Bara09}%
  \BibitemOpen
  \bibfield  {author} {\bibinfo {author} {\bibfnamefont {L.}~\bibnamefont {Barack}},\ }\href {\doibase 10.1088/0264-9381/26/21/213001} {\bibfield  {journal} {\bibinfo  {journal} {Class. Quant. Grav.}\ }\textbf {\bibinfo {volume} {26}},\ \bibinfo {pages} {213001} (\bibinfo {year} {2009})},\ \Eprint {http://arxiv.org/abs/0908.1664} {arXiv:0908.1664 [gr-qc]} \BibitemShut {NoStop}%
\bibitem [{\citenamefont {Poisson}\ \emph {et~al.}(2011)\citenamefont {Poisson}, \citenamefont {Pound},\ and\ \citenamefont {Vega}}]{PoisPounVega11}%
  \BibitemOpen
  \bibfield  {author} {\bibinfo {author} {\bibfnamefont {E.}~\bibnamefont {Poisson}}, \bibinfo {author} {\bibfnamefont {A.}~\bibnamefont {Pound}}, \ and\ \bibinfo {author} {\bibfnamefont {I.}~\bibnamefont {Vega}},\ }\href@noop {} {\bibfield  {journal} {\bibinfo  {journal} {Living Rev. Rel.}\ }\textbf {\bibinfo {volume} {14}},\ \bibinfo {pages} {7} (\bibinfo {year} {2011})},\ \Eprint {http://arxiv.org/abs/gr-qc/1102.0529} {arXiv:gr-qc/1102.0529} \BibitemShut {NoStop}%
\bibitem [{\citenamefont {Wardell}\ \emph {et~al.}(2023{\natexlab{a}})\citenamefont {Wardell}, \citenamefont {Pound}, \citenamefont {Warburton}, \citenamefont {Miller}, \citenamefont {Durkan},\ and\ \citenamefont {Le~Tiec}}]{WardETC23}%
  \BibitemOpen
  \bibfield  {author} {\bibinfo {author} {\bibfnamefont {B.}~\bibnamefont {Wardell}}, \bibinfo {author} {\bibfnamefont {A.}~\bibnamefont {Pound}}, \bibinfo {author} {\bibfnamefont {N.}~\bibnamefont {Warburton}}, \bibinfo {author} {\bibfnamefont {J.}~\bibnamefont {Miller}}, \bibinfo {author} {\bibfnamefont {L.}~\bibnamefont {Durkan}}, \ and\ \bibinfo {author} {\bibfnamefont {A.}~\bibnamefont {Le~Tiec}},\ }\href {\doibase 10.1103/PhysRevLett.130.241402} {\bibfield  {journal} {\bibinfo  {journal} {Phys. Rev. Lett.}\ }\textbf {\bibinfo {volume} {130}},\ \bibinfo {pages} {241402} (\bibinfo {year} {2023}{\natexlab{a}})}\BibitemShut {NoStop}%
\bibitem [{\citenamefont {Burke}\ \emph {et~al.}(2024)\citenamefont {Burke}, \citenamefont {Piovano}, \citenamefont {Warburton}, \citenamefont {Lynch}, \citenamefont {Speri}, \citenamefont {Kavanagh}, \citenamefont {Wardell}, \citenamefont {Pound}, \citenamefont {Durkan},\ and\ \citenamefont {Miller}}]{Burke:2023lno}%
  \BibitemOpen
  \bibfield  {author} {\bibinfo {author} {\bibfnamefont {O.}~\bibnamefont {Burke}}, \bibinfo {author} {\bibfnamefont {G.~A.}\ \bibnamefont {Piovano}}, \bibinfo {author} {\bibfnamefont {N.}~\bibnamefont {Warburton}}, \bibinfo {author} {\bibfnamefont {P.}~\bibnamefont {Lynch}}, \bibinfo {author} {\bibfnamefont {L.}~\bibnamefont {Speri}}, \bibinfo {author} {\bibfnamefont {C.}~\bibnamefont {Kavanagh}}, \bibinfo {author} {\bibfnamefont {B.}~\bibnamefont {Wardell}}, \bibinfo {author} {\bibfnamefont {A.}~\bibnamefont {Pound}}, \bibinfo {author} {\bibfnamefont {L.}~\bibnamefont {Durkan}}, \ and\ \bibinfo {author} {\bibfnamefont {J.}~\bibnamefont {Miller}},\ }\href {\doibase 10.1103/PhysRevD.109.124048} {\bibfield  {journal} {\bibinfo  {journal} {Phys. Rev. D}\ }\textbf {\bibinfo {volume} {109}},\ \bibinfo {pages} {124048} (\bibinfo {year} {2024})},\ \Eprint {http://arxiv.org/abs/2310.08927} {arXiv:2310.08927 [gr-qc]} \BibitemShut {NoStop}%
\bibitem [{\citenamefont {{Peters}}(1964)}]{Pete64}%
  \BibitemOpen
  \bibfield  {author} {\bibinfo {author} {\bibfnamefont {P.~C.}\ \bibnamefont {{Peters}}},\ }\href {\doibase 10.1103/PhysRev.136.B1224} {\bibfield  {journal} {\bibinfo  {journal} {Physical Review}\ }\textbf {\bibinfo {volume} {136}},\ \bibinfo {pages} {B1224} (\bibinfo {year} {1964})}\BibitemShut {NoStop}%
\bibitem [{\citenamefont {Glampedakis}\ and\ \citenamefont {Kennefick}(2002)}]{GlamKenn02}%
  \BibitemOpen
  \bibfield  {author} {\bibinfo {author} {\bibfnamefont {K.}~\bibnamefont {Glampedakis}}\ and\ \bibinfo {author} {\bibfnamefont {D.~J.}\ \bibnamefont {Kennefick}},\ }\href@noop {} {\bibfield  {journal} {\bibinfo  {journal} {Phys. Rev. D}\ }\textbf {\bibinfo {volume} {66}},\ \bibinfo {pages} {044002} (\bibinfo {year} {2002})}\BibitemShut {NoStop}%
\bibitem [{\citenamefont {{Hughes}}(2000)}]{Hugh00b}%
  \BibitemOpen
  \bibfield  {author} {\bibinfo {author} {\bibfnamefont {S.~A.}\ \bibnamefont {{Hughes}}},\ }\href {\doibase 10.1103/PhysRevD.61.084004} {\bibfield  {journal} {\bibinfo  {journal} {Phys. Rev. D}\ }\textbf {\bibinfo {volume} {61}},\ \bibinfo {eid} {084004} (\bibinfo {year} {2000})},\ \Eprint {http://arxiv.org/abs/gr-qc/9910091} {gr-qc/9910091} \BibitemShut {NoStop}%
\bibitem [{\citenamefont {Della~Rocca}\ \emph {et~al.}(2024)\citenamefont {Della~Rocca}, \citenamefont {Barsanti}, \citenamefont {Gualtieri},\ and\ \citenamefont {Maselli}}]{DellETC24}%
  \BibitemOpen
  \bibfield  {author} {\bibinfo {author} {\bibfnamefont {M.}~\bibnamefont {Della~Rocca}}, \bibinfo {author} {\bibfnamefont {S.}~\bibnamefont {Barsanti}}, \bibinfo {author} {\bibfnamefont {L.}~\bibnamefont {Gualtieri}}, \ and\ \bibinfo {author} {\bibfnamefont {A.}~\bibnamefont {Maselli}},\ }\href@noop {} {\  (\bibinfo {year} {2024})},\ \Eprint {http://arxiv.org/abs/2401.09542} {arXiv:2401.09542 [gr-qc]} \BibitemShut {NoStop}%
\bibitem [{\citenamefont {{Mano}}\ \emph {et~al.}(1996{\natexlab{a}})\citenamefont {{Mano}}, \citenamefont {{Suzuki}},\ and\ \citenamefont {{Takasugi}}}]{ManoSuzuTaka96a}%
  \BibitemOpen
  \bibfield  {author} {\bibinfo {author} {\bibfnamefont {S.}~\bibnamefont {{Mano}}}, \bibinfo {author} {\bibfnamefont {H.}~\bibnamefont {{Suzuki}}}, \ and\ \bibinfo {author} {\bibfnamefont {E.}~\bibnamefont {{Takasugi}}},\ }\href {\doibase 10.1143/PTP.96.549} {\bibfield  {journal} {\bibinfo  {journal} {Progress of Theoretical Physics}\ }\textbf {\bibinfo {volume} {96}},\ \bibinfo {pages} {549} (\bibinfo {year} {1996}{\natexlab{a}})},\ \Eprint {http://arxiv.org/abs/gr-qc/9605057} {gr-qc/9605057} \BibitemShut {NoStop}%
\bibitem [{\citenamefont {{Mano}}\ \emph {et~al.}(1996{\natexlab{b}})\citenamefont {{Mano}}, \citenamefont {{Suzuki}},\ and\ \citenamefont {{Takasugi}}}]{ManoSuzuTaka96b}%
  \BibitemOpen
  \bibfield  {author} {\bibinfo {author} {\bibfnamefont {S.}~\bibnamefont {{Mano}}}, \bibinfo {author} {\bibfnamefont {H.}~\bibnamefont {{Suzuki}}}, \ and\ \bibinfo {author} {\bibfnamefont {E.}~\bibnamefont {{Takasugi}}},\ }\href {\doibase 10.1143/PTP.95.1079} {\bibfield  {journal} {\bibinfo  {journal} {Progress of Theoretical Physics}\ }\textbf {\bibinfo {volume} {95}},\ \bibinfo {pages} {1079} (\bibinfo {year} {1996}{\natexlab{b}})},\ \Eprint {http://arxiv.org/abs/gr-qc/9603020} {gr-qc/9603020} \BibitemShut {NoStop}%
\bibitem [{\citenamefont {Isoyama}\ \emph {et~al.}(2022)\citenamefont {Isoyama}, \citenamefont {Fujita}, \citenamefont {Chua}, \citenamefont {Nakano}, \citenamefont {Pound},\ and\ \citenamefont {Sago}}]{IsoyETC22}%
  \BibitemOpen
  \bibfield  {author} {\bibinfo {author} {\bibfnamefont {S.}~\bibnamefont {Isoyama}}, \bibinfo {author} {\bibfnamefont {R.}~\bibnamefont {Fujita}}, \bibinfo {author} {\bibfnamefont {A.~J.~K.}\ \bibnamefont {Chua}}, \bibinfo {author} {\bibfnamefont {H.}~\bibnamefont {Nakano}}, \bibinfo {author} {\bibfnamefont {A.}~\bibnamefont {Pound}}, \ and\ \bibinfo {author} {\bibfnamefont {N.}~\bibnamefont {Sago}},\ }\href {\doibase 10.1103/PhysRevLett.128.231101} {\bibfield  {journal} {\bibinfo  {journal} {Phys. Rev. Lett.}\ }\textbf {\bibinfo {volume} {128}},\ \bibinfo {pages} {231101} (\bibinfo {year} {2022})}\BibitemShut {NoStop}%
\bibitem [{\citenamefont {{Kavanagh}}\ \emph {et~al.}(2016)\citenamefont {{Kavanagh}}, \citenamefont {{Ottewill}},\ and\ \citenamefont {{Wardell}}}]{KavaOtteWard16}%
  \BibitemOpen
  \bibfield  {author} {\bibinfo {author} {\bibfnamefont {C.}~\bibnamefont {{Kavanagh}}}, \bibinfo {author} {\bibfnamefont {A.~C.}\ \bibnamefont {{Ottewill}}}, \ and\ \bibinfo {author} {\bibfnamefont {B.}~\bibnamefont {{Wardell}}},\ }\href {\doibase 10.1103/PhysRevD.93.124038} {\bibfield  {journal} {\bibinfo  {journal} {Phys. Rev. D}\ }\textbf {\bibinfo {volume} {93}},\ \bibinfo {eid} {124038} (\bibinfo {year} {2016})},\ \Eprint {http://arxiv.org/abs/1601.03394} {arXiv:1601.03394 [gr-qc]} \BibitemShut {NoStop}%
\bibitem [{\citenamefont {Munna}(2023)}]{Munn23}%
  \BibitemOpen
  \bibfield  {author} {\bibinfo {author} {\bibfnamefont {C.}~\bibnamefont {Munna}},\ }\href {\doibase 10.1103/PhysRevD.108.084012} {\bibfield  {journal} {\bibinfo  {journal} {Phys. Rev. D}\ }\textbf {\bibinfo {volume} {108}},\ \bibinfo {pages} {084012} (\bibinfo {year} {2023})}\BibitemShut {NoStop}%
\bibitem [{\citenamefont {Warburton}\ \emph {et~al.}(2023)\citenamefont {Warburton}, \citenamefont {Wardell}, \citenamefont {Munna},\ and\ \citenamefont {Kavanagh}}]{PostNewtonianSelfForce}%
  \BibitemOpen
  \bibfield  {author} {\bibinfo {author} {\bibfnamefont {N.}~\bibnamefont {Warburton}}, \bibinfo {author} {\bibfnamefont {B.}~\bibnamefont {Wardell}}, \bibinfo {author} {\bibfnamefont {C.}~\bibnamefont {Munna}}, \ and\ \bibinfo {author} {\bibfnamefont {C.}~\bibnamefont {Kavanagh}},\ }\href {\doibase 10.5281/zenodo.8112975} {\enquote {\bibinfo {title} {Postnewtonianselfforce},}\ } (\bibinfo {year} {2023}),\ \bibinfo {note} {\href{https://doi.org/10.5281/zenodo.8112975}{https://doi.org/10.5281/zenodo.8112975}}\BibitemShut {NoStop}%
\bibitem [{BHP()}]{BHPTK18}%
  \BibitemOpen
  \href@noop {} {\enquote {\bibinfo {title} {{Black Hole Perturbation Toolkit}},}\ }\bibinfo {note} {\url{bhptoolkit.org}}\BibitemShut {NoStop}%
\bibitem [{UNC()}]{UNCGrav22}%
  \BibitemOpen
  \href@noop {} {\enquote {\bibinfo {title} {{UNC Gravitational Physics Group}},}\ }\bibinfo {note} {\url{https://github.com/UNC-Gravitational-Physics}}\BibitemShut {NoStop}%
\bibitem [{\citenamefont {Carter}(1968)}]{Cart68}%
  \BibitemOpen
  \bibfield  {author} {\bibinfo {author} {\bibfnamefont {B.}~\bibnamefont {Carter}},\ }\href {\doibase 10.1103/PhysRev.174.1559} {\bibfield  {journal} {\bibinfo  {journal} {Phys. Rev.}\ }\textbf {\bibinfo {volume} {174}},\ \bibinfo {pages} {1559} (\bibinfo {year} {1968})}\BibitemShut {NoStop}%
\bibitem [{\citenamefont {Schmidt}(2002)}]{Schm02}%
  \BibitemOpen
  \bibfield  {author} {\bibinfo {author} {\bibfnamefont {W.}~\bibnamefont {Schmidt}},\ }\href {\doibase 10.1088/0264-9381/19/10/314} {\bibfield  {journal} {\bibinfo  {journal} {Class. Quant. Grav.}\ }\textbf {\bibinfo {volume} {19}},\ \bibinfo {pages} {2743} (\bibinfo {year} {2002})},\ \Eprint {http://arxiv.org/abs/gr-qc/0202090} {arXiv:gr-qc/0202090} \BibitemShut {NoStop}%
\bibitem [{\citenamefont {{Drasco}}\ and\ \citenamefont {{Hughes}}(2006)}]{DrasHugh06}%
  \BibitemOpen
  \bibfield  {author} {\bibinfo {author} {\bibfnamefont {S.}~\bibnamefont {{Drasco}}}\ and\ \bibinfo {author} {\bibfnamefont {S.~A.}\ \bibnamefont {{Hughes}}},\ }\href {\doibase 10.1103/PhysRevD.73.024027} {\bibfield  {journal} {\bibinfo  {journal} {Phys. Rev. D}\ }\textbf {\bibinfo {volume} {73}},\ \bibinfo {eid} {024027} (\bibinfo {year} {2006})},\ \Eprint {http://arxiv.org/abs/gr-qc/0509101} {gr-qc/0509101} \BibitemShut {NoStop}%
\bibitem [{\citenamefont {Mino}(2003)}]{Mino03}%
  \BibitemOpen
  \bibfield  {author} {\bibinfo {author} {\bibfnamefont {Y.}~\bibnamefont {Mino}},\ }\href {\doibase 10.1103/PhysRevD.67.084027} {\bibfield  {journal} {\bibinfo  {journal} {Phys. Rev. D}\ }\textbf {\bibinfo {volume} {67}},\ \bibinfo {pages} {084027} (\bibinfo {year} {2003})},\ \Eprint {http://arxiv.org/abs/gr-qc/0302075} {arXiv:gr-qc/0302075} \BibitemShut {NoStop}%
\bibitem [{\citenamefont {{Fujita}}\ and\ \citenamefont {{Hikida}}(2009)}]{FujiHiki09}%
  \BibitemOpen
  \bibfield  {author} {\bibinfo {author} {\bibfnamefont {R.}~\bibnamefont {{Fujita}}}\ and\ \bibinfo {author} {\bibfnamefont {W.}~\bibnamefont {{Hikida}}},\ }\href {\doibase 10.1088/0264-9381/26/13/135002} {\bibfield  {journal} {\bibinfo  {journal} {Classical and Quantum Gravity}\ }\textbf {\bibinfo {volume} {26}},\ \bibinfo {eid} {135002} (\bibinfo {year} {2009})},\ \Eprint {http://arxiv.org/abs/0906.1420} {arXiv:0906.1420 [gr-qc]} \BibitemShut {NoStop}%
\bibitem [{\citenamefont {Hughes}(2001)}]{Hugh01}%
  \BibitemOpen
  \bibfield  {author} {\bibinfo {author} {\bibfnamefont {S.~A.}\ \bibnamefont {Hughes}},\ }\href {\doibase 10.1103/PhysRevD.64.064004} {\bibfield  {journal} {\bibinfo  {journal} {Phys. Rev. D}\ }\textbf {\bibinfo {volume} {64}},\ \bibinfo {pages} {064004} (\bibinfo {year} {2001})}\BibitemShut {NoStop}%
\bibitem [{\citenamefont {{Darwin}}(1959)}]{Darw59}%
  \BibitemOpen
  \bibfield  {author} {\bibinfo {author} {\bibfnamefont {C.}~\bibnamefont {{Darwin}}},\ }\href {\doibase 10.1098/rspa.1959.0015} {\bibfield  {journal} {\bibinfo  {journal} {Proc. R. Soc. Lond. A}\ }\textbf {\bibinfo {volume} {249}},\ \bibinfo {pages} {180} (\bibinfo {year} {1959})}\BibitemShut {NoStop}%
\bibitem [{\citenamefont {{Darwin}}(1961)}]{Darw61}%
  \BibitemOpen
  \bibfield  {author} {\bibinfo {author} {\bibfnamefont {C.}~\bibnamefont {{Darwin}}},\ }\href {\doibase 10.1098/rspa.1961.0142} {\bibfield  {journal} {\bibinfo  {journal} {Proceedings of the Royal Society of London Series A}\ }\textbf {\bibinfo {volume} {263}},\ \bibinfo {pages} {39} (\bibinfo {year} {1961})}\BibitemShut {NoStop}%
\bibitem [{\citenamefont {Drasco}\ and\ \citenamefont {Hughes}(2004)}]{DrasHugh04}%
  \BibitemOpen
  \bibfield  {author} {\bibinfo {author} {\bibfnamefont {S.}~\bibnamefont {Drasco}}\ and\ \bibinfo {author} {\bibfnamefont {S.~A.}\ \bibnamefont {Hughes}},\ }\href {\doibase 10.1103/PhysRevD.69.044015} {\bibfield  {journal} {\bibinfo  {journal} {Phys. Rev. D}\ }\textbf {\bibinfo {volume} {69}},\ \bibinfo {pages} {044015} (\bibinfo {year} {2004})}\BibitemShut {NoStop}%
\bibitem [{\citenamefont {{Warburton}}(2015)}]{Warb15}%
  \BibitemOpen
  \bibfield  {author} {\bibinfo {author} {\bibfnamefont {N.}~\bibnamefont {{Warburton}}},\ }\href {\doibase 10.1103/PhysRevD.91.024045} {\bibfield  {journal} {\bibinfo  {journal} {Phys. Rev. D}\ }\textbf {\bibinfo {volume} {91}},\ \bibinfo {eid} {024045} (\bibinfo {year} {2015})},\ \Eprint {http://arxiv.org/abs/1408.2885} {arXiv:1408.2885 [gr-qc]} \BibitemShut {NoStop}%
\bibitem [{\citenamefont {Teukolsky}(1972)}]{Teuk72}%
  \BibitemOpen
  \bibfield  {author} {\bibinfo {author} {\bibfnamefont {S.~A.}\ \bibnamefont {Teukolsky}},\ }\href {\doibase 10.1103/PhysRevLett.29.1114} {\bibfield  {journal} {\bibinfo  {journal} {Phys. Rev. Lett.}\ }\textbf {\bibinfo {volume} {29}},\ \bibinfo {pages} {1114} (\bibinfo {year} {1972})}\BibitemShut {NoStop}%
\bibitem [{\citenamefont {Teukolsky}(1973)}]{Teuk73}%
  \BibitemOpen
  \bibfield  {author} {\bibinfo {author} {\bibfnamefont {S.}~\bibnamefont {Teukolsky}},\ }\href@noop {} {\bibfield  {journal} {\bibinfo  {journal} {Astrophys. J.}\ }\textbf {\bibinfo {volume} {185}},\ \bibinfo {pages} {635} (\bibinfo {year} {1973})}\BibitemShut {NoStop}%
\bibitem [{\citenamefont {Wardell}\ \emph {et~al.}(2024{\natexlab{a}})\citenamefont {Wardell}, \citenamefont {Kavanagh},\ and\ \citenamefont {Dolan}}]{Wardell:2024yoi}%
  \BibitemOpen
  \bibfield  {author} {\bibinfo {author} {\bibfnamefont {B.}~\bibnamefont {Wardell}}, \bibinfo {author} {\bibfnamefont {C.}~\bibnamefont {Kavanagh}}, \ and\ \bibinfo {author} {\bibfnamefont {S.~R.}\ \bibnamefont {Dolan}},\ }\href@noop {} {\  (\bibinfo {year} {2024}{\natexlab{a}})},\ \Eprint {http://arxiv.org/abs/2406.12510} {arXiv:2406.12510 [gr-qc]} \BibitemShut {NoStop}%
\bibitem [{\citenamefont {{Abramowitz}}\ and\ \citenamefont {{Stegun}}(1972)}]{AbraSteg72}%
  \BibitemOpen
  \bibfield  {author} {\bibinfo {author} {\bibfnamefont {M.}~\bibnamefont {{Abramowitz}}}\ and\ \bibinfo {author} {\bibfnamefont {I.~A.}\ \bibnamefont {{Stegun}}},\ }\href@noop {} {\emph {\bibinfo {title} {Handbook of Mathematical Functions, New York: Dover, 1972}}}\ (\bibinfo {year} {1972})\BibitemShut {NoStop}%
\bibitem [{\citenamefont {{Sasaki}}\ and\ \citenamefont {{Tagoshi}}(2003)}]{SasaTago03}%
  \BibitemOpen
  \bibfield  {author} {\bibinfo {author} {\bibfnamefont {M.}~\bibnamefont {{Sasaki}}}\ and\ \bibinfo {author} {\bibfnamefont {H.}~\bibnamefont {{Tagoshi}}},\ }\href {\doibase 10.12942/lrr-2003-6} {\bibfield  {journal} {\bibinfo  {journal} {Living Reviews in Relativity}\ }\textbf {\bibinfo {volume} {6}},\ \bibinfo {pages} {6} (\bibinfo {year} {2003})},\ \Eprint {http://arxiv.org/abs/gr-qc/0306120} {gr-qc/0306120} \BibitemShut {NoStop}%
\bibitem [{\citenamefont {{Teukolsky}}\ and\ \citenamefont {{Press}}(1974)}]{TeukPres74}%
  \BibitemOpen
  \bibfield  {author} {\bibinfo {author} {\bibfnamefont {S.~A.}\ \bibnamefont {{Teukolsky}}}\ and\ \bibinfo {author} {\bibfnamefont {W.~H.}\ \bibnamefont {{Press}}},\ }\href {\doibase 10.1086/153180} {\bibfield  {journal} {\bibinfo  {journal} {Astrophys. J.}\ }\textbf {\bibinfo {volume} {193}},\ \bibinfo {pages} {443} (\bibinfo {year} {1974})}\BibitemShut {NoStop}%
\bibitem [{\citenamefont {Bautista}\ \emph {et~al.}(2024)\citenamefont {Bautista}, \citenamefont {Bonelli}, \citenamefont {Iossa}, \citenamefont {Tanzini},\ and\ \citenamefont {Zhou}}]{Bautista:2023sdf}%
  \BibitemOpen
  \bibfield  {author} {\bibinfo {author} {\bibfnamefont {Y.~F.}\ \bibnamefont {Bautista}}, \bibinfo {author} {\bibfnamefont {G.}~\bibnamefont {Bonelli}}, \bibinfo {author} {\bibfnamefont {C.}~\bibnamefont {Iossa}}, \bibinfo {author} {\bibfnamefont {A.}~\bibnamefont {Tanzini}}, \ and\ \bibinfo {author} {\bibfnamefont {Z.}~\bibnamefont {Zhou}},\ }\href {\doibase 10.1103/PhysRevD.109.084071} {\bibfield  {journal} {\bibinfo  {journal} {Phys. Rev. D}\ }\textbf {\bibinfo {volume} {109}},\ \bibinfo {pages} {084071} (\bibinfo {year} {2024})},\ \Eprint {http://arxiv.org/abs/2312.05965} {arXiv:2312.05965 [hep-th]} \BibitemShut {NoStop}%
\bibitem [{\citenamefont {Throwe}(2010)}]{Thro10}%
  \BibitemOpen
  \bibfield  {author} {\bibinfo {author} {\bibfnamefont {W.}~\bibnamefont {Throwe}},\ }\href@noop {} {\enquote {\bibinfo {title} {High precision calculation of generic extreme mass ratio inspirals},}\ } (\bibinfo {year} {2010}),\ \bibinfo {note} {http://hdl.handle.net/1721.1/61270}\BibitemShut {NoStop}%
\bibitem [{\citenamefont {Kavanagh}\ \emph {et~al.}(2015)\citenamefont {Kavanagh}, \citenamefont {Ottewill},\ and\ \citenamefont {Wardell}}]{KavaOtteWard15}%
  \BibitemOpen
  \bibfield  {author} {\bibinfo {author} {\bibfnamefont {C.}~\bibnamefont {Kavanagh}}, \bibinfo {author} {\bibfnamefont {A.~C.}\ \bibnamefont {Ottewill}}, \ and\ \bibinfo {author} {\bibfnamefont {B.}~\bibnamefont {Wardell}},\ }\href {\doibase 10.1103/PhysRevD.92.084025} {\bibfield  {journal} {\bibinfo  {journal} {Phys. Rev. D}\ }\textbf {\bibinfo {volume} {92}},\ \bibinfo {pages} {084025} (\bibinfo {year} {2015})},\ \Eprint {http://arxiv.org/abs/1503.02334} {arXiv:1503.02334 [gr-qc]} \BibitemShut {NoStop}%
\bibitem [{\citenamefont {Neef}\ \emph {et~al.}()\citenamefont {Neef}, \citenamefont {Ottewill},\ and\ \citenamefont {Kavanagh}}]{SFPN}%
  \BibitemOpen
  \bibfield  {author} {\bibinfo {author} {\bibfnamefont {J.}~\bibnamefont {Neef}}, \bibinfo {author} {\bibfnamefont {A.}~\bibnamefont {Ottewill}}, \ and\ \bibinfo {author} {\bibfnamefont {C.}~\bibnamefont {Kavanagh}},\ }\href@noop {} {}\bibinfo {note} {\href{https://gitlab.com/jakobneef/sfpn.git}{https://gitlab.com/jakobneef/sfpn.git}}\BibitemShut {NoStop}%
\bibitem [{\citenamefont {{Ganz}}\ \emph {et~al.}(2007)\citenamefont {{Ganz}}, \citenamefont {{Hikida}}, \citenamefont {{Nakano}}, \citenamefont {{Sago}},\ and\ \citenamefont {{Tanaka}}}]{GanzETC07}%
  \BibitemOpen
  \bibfield  {author} {\bibinfo {author} {\bibfnamefont {K.}~\bibnamefont {{Ganz}}}, \bibinfo {author} {\bibfnamefont {W.}~\bibnamefont {{Hikida}}}, \bibinfo {author} {\bibfnamefont {H.}~\bibnamefont {{Nakano}}}, \bibinfo {author} {\bibfnamefont {N.}~\bibnamefont {{Sago}}}, \ and\ \bibinfo {author} {\bibfnamefont {T.}~\bibnamefont {{Tanaka}}},\ }\href {\doibase 10.1143/PTP.117.1041} {\bibfield  {journal} {\bibinfo  {journal} {Progress of Theoretical Physics}\ }\textbf {\bibinfo {volume} {117}},\ \bibinfo {pages} {1041} (\bibinfo {year} {2007})},\ \Eprint {http://arxiv.org/abs/gr-qc/0702054} {gr-qc/0702054} \BibitemShut {NoStop}%
\bibitem [{\citenamefont {{Sago}}\ and\ \citenamefont {{Fujita}}(2015)}]{SagoFuji15}%
  \BibitemOpen
  \bibfield  {author} {\bibinfo {author} {\bibfnamefont {N.}~\bibnamefont {{Sago}}}\ and\ \bibinfo {author} {\bibfnamefont {R.}~\bibnamefont {{Fujita}}},\ }\href {\doibase 10.1093/ptep/ptv092} {\bibfield  {journal} {\bibinfo  {journal} {Progress of Theoretical and Experimental Physics}\ }\textbf {\bibinfo {volume} {2015}},\ \bibinfo {eid} {073E03} (\bibinfo {year} {2015})},\ \Eprint {http://arxiv.org/abs/1505.01600} {arXiv:1505.01600 [gr-qc]} \BibitemShut {NoStop}%
\bibitem [{\citenamefont {Fujita}\ and\ \citenamefont {Shibata}(2020)}]{FujiShib20}%
  \BibitemOpen
  \bibfield  {author} {\bibinfo {author} {\bibfnamefont {R.}~\bibnamefont {Fujita}}\ and\ \bibinfo {author} {\bibfnamefont {M.}~\bibnamefont {Shibata}},\ }\href {\doibase 10.1103/PhysRevD.102.064005} {\bibfield  {journal} {\bibinfo  {journal} {Phys. Rev. D}\ }\textbf {\bibinfo {volume} {102}},\ \bibinfo {pages} {064005} (\bibinfo {year} {2020})}\BibitemShut {NoStop}%
\bibitem [{\citenamefont {Wardell}\ \emph {et~al.}(2024{\natexlab{b}})\citenamefont {Wardell}, \citenamefont {Warburton}, \citenamefont {Fransen}, \citenamefont {Upton}, \citenamefont {Cunningham}, \citenamefont {Ottewill},\ and\ \citenamefont {Casals}}]{SpinWeightedSpheroidalHarmonics}%
  \BibitemOpen
  \bibfield  {author} {\bibinfo {author} {\bibfnamefont {B.}~\bibnamefont {Wardell}}, \bibinfo {author} {\bibfnamefont {N.}~\bibnamefont {Warburton}}, \bibinfo {author} {\bibfnamefont {K.}~\bibnamefont {Fransen}}, \bibinfo {author} {\bibfnamefont {S.~D.}\ \bibnamefont {Upton}}, \bibinfo {author} {\bibfnamefont {K.}~\bibnamefont {Cunningham}}, \bibinfo {author} {\bibfnamefont {A.}~\bibnamefont {Ottewill}}, \ and\ \bibinfo {author} {\bibfnamefont {M.}~\bibnamefont {Casals}},\ }\href {\doibase 10.5281/zenodo.11199019} {\enquote {\bibinfo {title} {{SpinWeightedSpheroidalHarmonics}},}\ } (\bibinfo {year} {2024}{\natexlab{b}}),\ \bibinfo {note} {\href{https://doi.org/10.5281/zenodo.11199019}{https://doi.org/10.5281/zenodo.11199019}}\BibitemShut {NoStop}%
\bibitem [{\citenamefont {{Munna}}(2020)}]{Munn20}%
  \BibitemOpen
  \bibfield  {author} {\bibinfo {author} {\bibfnamefont {C.}~\bibnamefont {{Munna}}},\ }\href {\doibase 10.1103/PhysRevD.102.124001} {\bibfield  {journal} {\bibinfo  {journal} {Phys. Rev. D}\ }\textbf {\bibinfo {volume} {102}},\ \bibinfo {eid} {124001} (\bibinfo {year} {2020})},\ \Eprint {http://arxiv.org/abs/2008.10622} {arXiv:2008.10622 [gr-qc]} \BibitemShut {NoStop}%
\bibitem [{\citenamefont {{Blanchet}}(2014)}]{Blan14}%
  \BibitemOpen
  \bibfield  {author} {\bibinfo {author} {\bibfnamefont {L.}~\bibnamefont {{Blanchet}}},\ }\href {\doibase 10.12942/lrr-2014-2} {\bibfield  {journal} {\bibinfo  {journal} {Living Reviews in Relativity}\ }\textbf {\bibinfo {volume} {17}},\ \bibinfo {pages} {2} (\bibinfo {year} {2014})},\ \Eprint {http://arxiv.org/abs/1310.1528} {arXiv:1310.1528 [gr-qc]} \BibitemShut {NoStop}%
\bibitem [{\citenamefont {{Fujita}}(2012)}]{Fuji12b}%
  \BibitemOpen
  \bibfield  {author} {\bibinfo {author} {\bibfnamefont {R.}~\bibnamefont {{Fujita}}},\ }\href@noop {} {\bibfield  {journal} {\bibinfo  {journal} {Progress of Theoretical Physics}\ }\textbf {\bibinfo {volume} {128}},\ \bibinfo {pages} {971} (\bibinfo {year} {2012})},\ \Eprint {http://arxiv.org/abs/1211.5535} {arXiv:1211.5535 [gr-qc]} \BibitemShut {NoStop}%
\bibitem [{\citenamefont {{Fujita}}(2015)}]{Fuji15}%
  \BibitemOpen
  \bibfield  {author} {\bibinfo {author} {\bibfnamefont {R.}~\bibnamefont {{Fujita}}},\ }\href {\doibase 10.1093/ptep/ptv012} {\bibfield  {journal} {\bibinfo  {journal} {Prog. Theor. Exp. Phys.}\ }\textbf {\bibinfo {volume} {2015}} (\bibinfo {year} {2015}),\ 10.1093/ptep/ptv012},\ \Eprint {http://arxiv.org/abs/1412.5689} {arXiv:1412.5689 [gr-qc]} \BibitemShut {NoStop}%
\bibitem [{{\relax DLMF}()}]{DLMF}%
  \BibitemOpen
  {\relax DLMF},\ \href {http://dlmf.nist.gov/} {\enquote {\bibinfo {title} {{NIST Digital Library of Mathematical Functions}},}\ }\bibinfo {howpublished} {http://dlmf.nist.gov/, Release 1.0.10 of 2015-08-07},\ \bibinfo {note} {online companion to \cite{NHMF}}\BibitemShut {NoStop}%
\bibitem [{\citenamefont {{Taracchini}}\ \emph {et~al.}(2013)\citenamefont {{Taracchini}}, \citenamefont {{Buonanno}}, \citenamefont {{Hughes}},\ and\ \citenamefont {{Khanna}}}]{TaraETC13}%
  \BibitemOpen
  \bibfield  {author} {\bibinfo {author} {\bibfnamefont {A.}~\bibnamefont {{Taracchini}}}, \bibinfo {author} {\bibfnamefont {A.}~\bibnamefont {{Buonanno}}}, \bibinfo {author} {\bibfnamefont {S.}~\bibnamefont {{Hughes}}}, \ and\ \bibinfo {author} {\bibfnamefont {G.}~\bibnamefont {{Khanna}}},\ }\href {\doibase 10.1103/PhysRevD.88.0440012} {\bibfield  {journal} {\bibinfo  {journal} {Phys. Rev. D}\ }\textbf {\bibinfo {volume} {88}},\ \bibinfo {eid} {044001} (\bibinfo {year} {2013})},\ \Eprint {http://arxiv.org/abs/1305.2184} {arXiv:1305.2184 [gr-qc]} \BibitemShut {NoStop}%
\bibitem [{\citenamefont {Shah}(2014)}]{Shah14}%
  \BibitemOpen
  \bibfield  {author} {\bibinfo {author} {\bibfnamefont {A.~G.}\ \bibnamefont {Shah}},\ }\href {\doibase 10.1103/PhysRevD.90.044025} {\bibfield  {journal} {\bibinfo  {journal} {Phys. Rev. D}\ }\textbf {\bibinfo {volume} {90}},\ \bibinfo {pages} {044025} (\bibinfo {year} {2014})},\ \Eprint {http://arxiv.org/abs/1403.2697} {arXiv:1403.2697 [gr-qc]} \BibitemShut {NoStop}%
\bibitem [{\citenamefont {{Johnson-McDaniel}}\ \emph {et~al.}(2015)\citenamefont {{Johnson-McDaniel}}, \citenamefont {{Shah}},\ and\ \citenamefont {{Whiting}}}]{JohnMcDaShahWhit15}%
  \BibitemOpen
  \bibfield  {author} {\bibinfo {author} {\bibfnamefont {N.~K.}\ \bibnamefont {{Johnson-McDaniel}}}, \bibinfo {author} {\bibfnamefont {A.~G.}\ \bibnamefont {{Shah}}}, \ and\ \bibinfo {author} {\bibfnamefont {B.~F.}\ \bibnamefont {{Whiting}}},\ }\href {\doibase 10.1103/PhysRevD.92.044007} {\bibfield  {journal} {\bibinfo  {journal} {Phys. Rev. D}\ }\textbf {\bibinfo {volume} {92}},\ \bibinfo {eid} {044007} (\bibinfo {year} {2015})},\ \Eprint {http://arxiv.org/abs/1503.02638} {arXiv:1503.02638 [gr-qc]} \BibitemShut {NoStop}%
\bibitem [{\citenamefont {{Munna}}\ and\ \citenamefont {{Evans}}(2019)}]{MunnEvan19a}%
  \BibitemOpen
  \bibfield  {author} {\bibinfo {author} {\bibfnamefont {C.}~\bibnamefont {{Munna}}}\ and\ \bibinfo {author} {\bibfnamefont {C.~R.}\ \bibnamefont {{Evans}}},\ }\href {\doibase 10.1103/PhysRevD.100.104060} {\bibfield  {journal} {\bibinfo  {journal} {Phys. Rev. D}\ }\textbf {\bibinfo {volume} {100}},\ \bibinfo {eid} {104060} (\bibinfo {year} {2019})},\ \Eprint {http://arxiv.org/abs/1909.05877} {arXiv:1909.05877 [gr-qc]} \BibitemShut {NoStop}%
\bibitem [{\citenamefont {Munna}\ and\ \citenamefont {Evans}(2020)}]{MunnEvan20a}%
  \BibitemOpen
  \bibfield  {author} {\bibinfo {author} {\bibfnamefont {C.}~\bibnamefont {Munna}}\ and\ \bibinfo {author} {\bibfnamefont {C.~R.}\ \bibnamefont {Evans}},\ }\href {\doibase 10.1103/PhysRevD.102.104006} {\bibfield  {journal} {\bibinfo  {journal} {Phys. Rev. D}\ }\textbf {\bibinfo {volume} {102}},\ \bibinfo {eid} {104006} (\bibinfo {year} {2020})},\ \Eprint {http://arxiv.org/abs/2009.01254} {arXiv:2009.01254 [gr-qc]} \BibitemShut {NoStop}%
\bibitem [{\citenamefont {Wardell}\ \emph {et~al.}(2023{\natexlab{b}})\citenamefont {Wardell}, \citenamefont {Warburton}, \citenamefont {Cunningham}, \citenamefont {Durkan}, \citenamefont {Leather}, \citenamefont {Nasipak}, \citenamefont {Kavanagh}, \citenamefont {Torres}, \citenamefont {Ottewill},\ and\ \citenamefont {Casals}}]{TeukolskyBHPT}%
  \BibitemOpen
  \bibfield  {author} {\bibinfo {author} {\bibfnamefont {B.}~\bibnamefont {Wardell}}, \bibinfo {author} {\bibfnamefont {N.}~\bibnamefont {Warburton}}, \bibinfo {author} {\bibfnamefont {K.}~\bibnamefont {Cunningham}}, \bibinfo {author} {\bibfnamefont {L.}~\bibnamefont {Durkan}}, \bibinfo {author} {\bibfnamefont {B.}~\bibnamefont {Leather}}, \bibinfo {author} {\bibfnamefont {Z.}~\bibnamefont {Nasipak}}, \bibinfo {author} {\bibfnamefont {C.}~\bibnamefont {Kavanagh}}, \bibinfo {author} {\bibfnamefont {T.}~\bibnamefont {Torres}}, \bibinfo {author} {\bibfnamefont {A.}~\bibnamefont {Ottewill}}, \ and\ \bibinfo {author} {\bibfnamefont {M.}~\bibnamefont {Casals}},\ }\href {\doibase 10.5281/zenodo.10040501} {\enquote {\bibinfo {title} {Teukolsky},}\ } (\bibinfo {year} {2023}{\natexlab{b}}),\ \bibinfo {note} {\href{https://doi.org/10.5281/zenodo.10040501}{https://doi.org/10.5281/zenodo.10040501}}\BibitemShut {NoStop}%
\bibitem [{\citenamefont {{Sago}}\ \emph {et~al.}(2024)\citenamefont {{Sago}}, \citenamefont {{Fujita}},\ and\ \citenamefont {{Nakano}}}]{SagoFujiNaka24}%
  \BibitemOpen
  \bibfield  {author} {\bibinfo {author} {\bibfnamefont {N.}~\bibnamefont {{Sago}}}, \bibinfo {author} {\bibfnamefont {R.}~\bibnamefont {{Fujita}}}, \ and\ \bibinfo {author} {\bibfnamefont {H.}~\bibnamefont {{Nakano}}},\ }\href {\doibase 10.48550/arXiv.2411.09147} {\bibfield  {journal} {\bibinfo  {journal} {arXiv e-prints}\ ,\ \bibinfo {eid} {arXiv:2411.09147}} (\bibinfo {year} {2024})},\ \Eprint {http://arxiv.org/abs/2411.09147} {arXiv:2411.09147 [gr-qc]} \BibitemShut {NoStop}%
\bibitem [{\citenamefont {Stein}\ and\ \citenamefont {Warburton}(2020)}]{SteiWarb19}%
  \BibitemOpen
  \bibfield  {author} {\bibinfo {author} {\bibfnamefont {L.~C.}\ \bibnamefont {Stein}}\ and\ \bibinfo {author} {\bibfnamefont {N.}~\bibnamefont {Warburton}},\ }\href {\doibase 10.1103/PhysRevD.101.064007} {\bibfield  {journal} {\bibinfo  {journal} {Phys. Rev. D}\ }\textbf {\bibinfo {volume} {101}},\ \bibinfo {pages} {064007} (\bibinfo {year} {2020})}\BibitemShut {NoStop}%
\bibitem [{\citenamefont {Olver}\ \emph {et~al.}(2010)\citenamefont {Olver}, \citenamefont {Lozier}, \citenamefont {Boisvert},\ and\ \citenamefont {Clark}}]{NHMF}%
  \BibitemOpen
  \bibinfo {editor} {\bibfnamefont {F.~W.~J.}\ \bibnamefont {Olver}}, \bibinfo {editor} {\bibfnamefont {D.~W.}\ \bibnamefont {Lozier}}, \bibinfo {editor} {\bibfnamefont {R.~F.}\ \bibnamefont {Boisvert}}, \ and\ \bibinfo {editor} {\bibfnamefont {C.~W.}\ \bibnamefont {Clark}},\ eds.,\ \href@noop {} {\emph {\bibinfo {title} {{NIST Handbook of Mathematical Functions}}}}\ (\bibinfo  {publisher} {Cambridge University Press},\ \bibinfo {address} {New York, NY},\ \bibinfo {year} {2010})\ \bibinfo {note} {print companion to \cite{DLMF}}\BibitemShut {NoStop}%
\end{thebibliography}%
\end{document}